\documentclass[12pt]{article}

\pdfoutput=1

\usepackage{amssymb}
\usepackage{amsmath}
\usepackage{graphicx}
\usepackage{amsfonts}
\usepackage{amssymb}

\def\del{\partial}

\def\half{\frac{1}{2}}

\makeatletter
\renewcommand\section{\@startsection {section}{1}{\z@}%
                                 {-3.5ex \@plus -1ex \@minus -.2ex}
                                   {2.3ex \@plus.2ex}%
                                   {\normalfont\large\bfseries}}
\renewcommand\subsection{\@startsection{subsection}{2}{\z@}%
                                   {-3.25ex\@plus -1ex \@minus -.2ex}%
                                     {1.5ex \@plus .2ex}%
                                     {\normalfont\bfseries}}
\renewcommand\subsubsection{\@startsection{subsubsection}{3}{\z@}%
                                   {-3.25ex\@plus -1ex \@minus -.2ex}%
                                     {1.5ex \@plus .2ex}%
                                     {\normalfont\itshape}}
\makeatother

\newcommand{\be}{\begin{equation}}
\newcommand{\ee}{\end{equation}}
\newcommand{\bea}{\begin{eqnarray}}
\newcommand{\eea}{\end{eqnarray}}
\newcommand{\barr}{\begin{array}}
\newcommand{\earr}{\end{array}}

\def\beq{\begin{equation}}
\def\eeq{\end{equation}}
\def\be{\begin{equation}}
\def\ee{\end{equation}}
\def\bea{\begin{eqnarray}}
\def\eea{\end{eqnarray}}

\DeclareRobustCommand{\SkipTocEntry}[4]{}

\begin{document}

\begin{titlepage}

\setcounter{page}{1} \baselineskip=15.5pt \thispagestyle{empty}

\begin{flushright}
\end{flushright}
\vfil

\begin{center}

{\LARGE  Gravity Waves \\ and \\ Linear Inflation from  Axion
Monodromy
\\}
%
%
%
%
%


\end{center}
\bigskip\

\begin{center}
{\large Liam McAllister,$^{\rm 1}$ Eva Silverstein,$^{\rm 2}$ and
Alexander Westphal$^{\rm 2}$}
\end{center}

\begin{center}
\textit{$^{\rm 1}$LEPP and Department of Physics, Cornell
University, Ithaca NY 14853}

\textit{$^{\rm 2}$SLAC and Department of Physics, Stanford
University, Stanford CA 94305}
\end{center} \vfil

\noindent  Wrapped branes in string compactifications introduce a
monodromy that extends the field range of individual closed-string
axions to beyond the Planck scale.  Furthermore, approximate shift
symmetries of the system naturally
control corrections to the axion potential.  This suggests a general
mechanism for chaotic inflation driven by monodromy-extended
closed-string axions.  We systematically analyze this possibility
and show that the mechanism is compatible with moduli stabilization
and can be realized in many types of compactifications, including
warped Calabi-Yau manifolds and more general Ricci-curved spaces.
In this broad class of models, the potential is linear in the
canonical inflaton field, predicting a tensor to scalar ratio
$r\approx 0.07$ accessible to upcoming cosmic microwave background
(CMB) observations.

\vfil
\begin{flushleft}
\today
\end{flushleft}

\end{titlepage}

\newpage
\tableofcontents
\newpage

\section{Introduction:  Axion Recycling}

An important class of inflationary models \cite{Inflation},  chaotic
inflation \cite{Linde:1983gd}, involves an inflaton field excursion
that is large compared to the Planck scale $M_P$ \cite{Lyth}.  These
models have a GUT-scale inflaton potential, and are accessible to
observational tests via a B-mode polarization signature in the CMB
\cite{Bmodes, Bmodeobs}.

The Planckian or super-Planckian field excursions required for
high-scale inflation may be formally protected by an approximate
shift symmetry in effective field theory. A canonical class of
examples with a field excursion $\Delta\Phi\simeq M_P$, known as
Natural Inflation, employs a pseudo-Nambu-Goldstone boson mode (an
axion) as the inflaton \cite{Freese:1990rb,OtherNatural}.

Because inflation is sensitive to Planck-suppressed operators,
however, it is still of significant interest to go beyond effective
field theory and realize inflation in string theory, a candidate
ultraviolet completion of gravity. Conversely, CMB observations
which discriminate among different inflationary mechanisms provide
an opportunity to probe some basic features of the ultraviolet
completion of gravity.

The lightest scalar fields in string compactifications roughly
divide into radial and angular moduli. Radial moduli, such as the
dilaton and the compactification volume, have an unbounded field
range as they go toward weak-coupling limits.  In these limits their
contributions to the potential are typically very steep, not
sourcing large-field inflation in any example yet studied. Angular
moduli, such as axions, have potentials that are classically
protected by shift symmetries. However, in the case of axions it has
been argued that the field range contained within a single period is
generally sub-Planckian in string theory \cite{Banks:2003sx},
leading to proposals to extend the field range by combining many
axions \cite{Nflation}.

In the present work, we show that in the presence of suitable
wrapped branes, the potential energy is no longer a periodic
function of the axion.  When this {\it monodromy} in the moduli
space is taken into account, a single axion develops a kinematically
unbounded field range with a potential energy growing {\it linearly}
with the canonically normalized inflaton field.  This implements the
monodromy mechanism introduced in \cite{Silverstein:2008sg}\ in a
wide class of string compactifications.

Because the basic idea is very simple, let us indicate it here.
Axions arise in string compactifications from integrating gauge
potentials over nontrivial cycles.  For example, in type IIB string
theory, there are axions $b_I=\int_{\Sigma_I^{(2)}}B$ arising from
integrating the  Neveu-Schwarz (NS) two-form potential $B_{MN}$ over
two-cycles $\Sigma_I^{(2)}$, and similarly axions
$c_I=\int_{\Sigma_I^{(2)}}C$ arise from the Ramond-Ramond (RR)
two-form $C_{MN}$.
In the absence of additional ingredients such as fluxes and
space-filling wrapped branes, the potential for these axions is
classically flat, and develops a periodic contribution from
instanton effects.  A Dp-brane wrapping $\Sigma_I^{(2)}$, on the
other hand, carries a potential energy that is {\it not} a periodic
function of the axion: in fact, this energy increases without bound
as $b_I$ increases.  The effective action for such a wrapped brane
is the DBI action, given in terms of the embedding coordinates
$X^M(\xi)$ as
s%
%
\be S_{DBI}=-\int
{d^{p+1}\xi\over{(2\pi)^p}}\alpha'^{-(p+1)/2}e^{-\Phi}\sqrt{\det{(G_{MN}+B_{MN})\,
\del_\alpha X^M\del_\beta X^N}} \label{DBIgen} \ee
where we have omitted the corresponding Chern-Simons term, which
will be unimportant for our considerations. A key example is a
D5-brane wrapped on a two-cycle $\Sigma^{(2)}$ of size
$\ell\sqrt{\alpha'}$, which yields a potential
\be V(b) = {\epsilon\over{g_s(2\pi)^5\alpha'^2}}\sqrt{\ell^4+b^2}
\label{Bpot} \ee
that is linear in the axion field $b$ at large $b$. (Here we have
included a factor $\epsilon$ to represent the effects of warping,
which we describe more carefully below.) Similarly, an NS5-brane
wrapped on $\Sigma_I^{(2)}$ introduces a monodromy in the $c_I$
direction.

Monodromy is a common phenomenon in string compactifications.  In
the past, it has been studied extensively in the context of particle
states in field theory \cite{Witten:1979ey}\ and the corresponding
non-space-filling wrapped branes of string theory
\cite{wrappedmonod}. The present case of monodromy in the {\it
potential energy} arises when a would-be periodic direction $\gamma$
is ``unwrapped'' by the inclusion of an additional space-filling ingredient
%
whose potential energy grows as one moves in the $\gamma$ direction,
extending the kinematic range of the corresponding scalar field.
Because the wrapped branes are space-filling, their charge must be
cancelled within the compactification.  We will do so with an
antibrane wrapped on a distant, homologous two-cycle as depicted in
Fig. 2 in \S4\ below.

In the bulk of this paper, we analyze the conditions under which
this yields controlled large-field inflation in string theory. We
find a reasonably natural class of viable models. As is usually the
case in
inflationary model building from string theory, much of the
challenge is to gain systematic control of Planck-suppressed
corrections to the
effective action.  After ensuring that our candidate inflaton
potential does not destabilize the compactification moduli, and that
fluxes do not affect the structure of our candidate inflaton
potential, we establish that instanton effects, which produce
sinusoidal contributions to the axion potential, can be naturally
suppressed.
%
We assess these conditions for both perturbative and nonperturbative
stabilization mechanisms, drawing examples based both on Calabi-Yau
compactifications and on more general compactifications that break
supersymmetry at the Kaluza-Klein scale.  In the case of
nonperturbative stabilization mechanisms in type IIB string theory,
we find a controlled set of models for the RR two-form axions $c_I$,
while perturbative stabilization mechanisms suggest opportunities
for inflating in the $b_I$ as well as in the $c_I$ directions.
These varied implementations of our axion monodromy mechanism give
identical predictions for the overall tilt and tensor to scalar
ratio in the CMB, as they are all well-described by a linear
potential for a canonically-normalized inflaton.\footnote{There may
also be novel signatures from finer details of the power spectrum
originating in the repeated circuits of the fundamental axion
period, as we discuss further below.} Our prediction for these
quantities lies well within the exclusion contours from present data
\cite{Observations}, and is ultimately distinguishable from the predictions of other
canonical models via planned CMB experiments \cite{Bmodeobs,Planck}
(see Fig.~3).

Our mechanism relies on specific additional ingredients -- branes --
intrinsic in the ultraviolet completion of gravity afforded by
string theory.  Although string theory restricts the range of the
original axion period in the first place, it then recycles a single
axion via monodromy, providing a simple generalization of
\cite{Linde:1983gd,Freese:1990rb}\ with its own distinctive
predictions.  The subject of axion inflation has thus almost come
full circle.\footnote{Though we hope to have added something to the
subject this time around.}


\section{Axions and the Candidate Inflaton Action}

Axions in string theory arise from integrating gauge potentials over
nontrivial cycles in the compactification manifold $X$. Let
$\Sigma_I$, $I=1,\ldots h^{1,1}(X)$  be an integral basis of
$H_2(X,\mathbb{Z})$, and let $\omega^I$ be a dual basis of
$H^2(X,\mathbb{Z})$,  with
$\int_{\Sigma_I}\omega^J=\alpha'\delta_I^{~J}$.  Then for the
Neveu-Schwarz two-form potential $B^{(2)}$, let us write
%
\be B^{(2)}= b_I(x)\omega_2^I \ee with $x$ the four-dimensional
spacetime coordinate.
%

In the case of type II theories, additional axions arise from
integrating the RR p-form potentials over p-cycles. Taking
$\omega^\alpha$, $\alpha=1,\ldots b^p(X)$,  to be a basis of
$H^p(X,\mathbb{Z})$ dual to an integral homology basis, we can write
\be C^{(p)}=c^{(p)}_{\alpha}(x)\omega_p^{\alpha} \ee
In type IIB string theory, for example, we have an RR two-form
$C^{(2)}$ which will play a key role in the case of Calabi-Yau
compactifications.

The period of these axions, collectively denoted by $a=\{b~{\rm
or}~c\}$, is
\be\label{axperiod} a\to a+(2\pi)^2 \ee
as can be seen from the worldsheet coupling
$(i/2\pi\alpha')\int_{\Sigma_I^{(2)}}B$ in the case of $B^{(2)}$.

%
%

%
%

\subsection{Axion Kinetic Terms}

In order to analyze the possibility of inflation with axions, we
will need their
kinetic and potential terms.
The classical kinetic term\footnote{The kinetic terms are in general
corrected by worldsheet instantons or D-instantons, in the cases of
$b$ and $c$, respectively. In our examples below we will ensure that
these instanton effects are negligible in our inflationary
solutions.}
 for the $b_I$ fields descends from the $|H_3|^2$ term in the ten-dimensional action, with $H_3=dB$.
In terms of the metric
\be \label{ds} ds^2 = g_{\mu\nu}dx^{\mu}dx^{\nu}+g_{ij}dy^i dy^j
\ee
we have \be \int d^{10}x {\sqrt{g}\over {(2\pi)^7
g_s^2\alpha'^4}}{1\over 2} |H|^2\Rightarrow {\cal S}_{kin,b}= \int{{
d^{10} x}\over{12(2\pi)^7
g_s^2\alpha'^4}}\sqrt{g}g^{\mu\nu}\del_{\mu}b_I\del_{\nu} b_J
\omega^I_{ij}\omega^J_{i'j'}g^{ii'}g^{jj'} \label{bkin}\ee
and similarly for the $C^{(p)}$ fields, with $F^{(p+1)}=dC^{(p)}$:
\be \int d^{10}x {\sqrt{g}\over {(2\pi)^7 \alpha'^4}} {1\over 2}
|F^{(p+1)}|^2\Rightarrow {\cal S}_{kin,c}= \int{{ d^{10}
x}\over{2(2\pi)^7 (p+1)!
\alpha'^4}}\sqrt{g}g^{\mu\nu}\del_{\mu}c_I\del_{\nu} c_J
\omega^I_{i_1\dots i_p}\omega^J_{i_1'\dots i_p'}g^{i_1i_1'}\dots
g^{i_p i_p'} \label{Ckin}\ee
To simplify the presentation we will now restrict attention to $b_I$
and to $c^{(2)}_{\alpha} \equiv c_I$, but the extension to other
$c^{(p)}_{\alpha}$ is immediate. The four-dimensional kinetic terms
for our axions $b_I, c_I$, collectively denoted as $a_I=\{b_I ~{\rm
or}~c_I\}$, may then be written
\be\label{genkin} S_{kin}=\half\int d^4
x\,\sqrt{g_4}~\gamma^{IJ}\,g^{\mu\nu}\del_{\mu}a_I\del_{\nu} a_J
\equiv\half\int d^4 x\,\sqrt{g_4}~\sum_I f_{a_I}^2\,(\del
a^{\prime}_I)^2 \equiv \half\int d^4 x\,\sqrt{g_4}~\sum_I (\del
\phi_{a_I})^2 \ee
%
where in the  second equality we have diagonalized the metric
$\gamma^{IJ}$, and in the third equality we have defined the
canonically-normalized axion field $\phi_{a_I}$ for the $I$th axion
of type $a=\{b~{\rm or}~ c\}$. In much of this paper, we will focus
on a single axion at a time, and use the notation $\phi_a$ for its
canonically normalized field. The canonically normalized inflaton
field has periodicity
\be\label{canonicalperiod} \phi_a\to\phi_a+(2\pi)^2 f_a \ee
corresponding to (\ref{axperiod}).

Using (\ref{bkin}), (\ref{Ckin}), the axion kinetic term depends on
the geometry of the compactification via \be \label{fbis}
\gamma^{IJ}= {\frac{1}{{6(2\pi)^7 g_s^2 \alpha'^4}}}\int
\omega_I\wedge\star\,\omega_J \ee for $b_I$, and \be \label{fcis}
\gamma^{IJ}= {\frac{1}{{6 (2\pi)^7 \alpha'^4}}} \int
\omega_I\wedge\star\,\omega_J \ee for $c_I$. To express these
results in terms of the four-dimensional reduced Planck mass
$M_{\rm P}$, we use
\be\label{alphapr} \alpha' M_{\rm P}^{2}=
\frac{2}{(2\pi)^7}\frac{\cal V}{g_s^2} \ee where ${\cal V}\alpha'^3$
is the volume of the compactification.

We will use (\ref{fbis}), (\ref{fcis}), guided by
\cite{Banks:2003sx,Svrcek:2006yi},  to determine the decay constants
in our specific examples below.  To provide intuition, we now record
the result in the simplified case in which all length scales
$L\sqrt{\alpha'}$ in the compactification are the same (and ${\cal
V}\equiv  L^6$).
%
From (\ref{bkin}) and (\ref{Ckin})\  we obtain
\be \label{samescales}\phi_b^2 \sim {L^2\over{3
g_s^2(2\pi)^{7}\alpha'}}b^2, ~~~~~~~~~ \phi_c^2\sim
{L^{6-2p}\over{3(2\pi)^{7}\alpha'}}c^2 ~~~~ {\rm (one~scale)} \ee
Using (\ref{alphapr})\ this gives
\be \label{samescalesMp} {\phi_b^2\over M_P^2}\sim {b^2\over{6
L^4}}, ~~~~~~~ {\phi_c^2\over M_P^2}\sim \frac{g_s^2 c^2}{6 L^4}
~~~~~ {\rm (one~scale)} \ee

\subsection{Wrapped Fivebrane Action}

As discussed in the introduction, wrapping appropriate branes on
cycles threaded by $B^{(2)}$ and $C^{(p)}$ introduces a non-periodic
potential for the axions $b$ and $c$. This follows immediately from
the DBI action (\ref{DBIgen}) in the case of D-branes on cycles with
$B$ fields, and can be seen by duality to apply to (p,q) fivebranes
on cycles with both $B$ and $C$ fields.

For D5-branes on a two-cycle $\Sigma^{(2)}$ of size $\ell
\sqrt{\alpha'}$ with $b$ axions turned on, or NS5-branes on a
two-cycle with a $c$ axion, we have
\be\label{basicpot}
V(b)={\epsilon\over{g_s(2\pi)^5\alpha'^2}}\sqrt{\ell^4+b^2} ~~~~~~~
V(c)= {\epsilon\over{g_s^2(2\pi)^5\alpha'^{2}}}\sqrt{\ell^4+c^2
g_s^2} \ee
where $\epsilon$ encodes warp-factor dependence to be discussed in
\S4.  A similar contribution arises from an anti-fivebrane wrapped
on a distant, homologous two-cycle as depicted in Fig. 2 below.

In the large-field regime of interest, this potential is linear in
the axion $a$, and hence in the canonically normalized field
$\phi_a$:
\be\label{genpot}V(\phi_a)\approx \mu_a^3\phi_a\ee
with $\mu_a$ a function of the parameters of the compactification
that depends on the model.  We will analyze its structure in detail
in several specific models in \S4 and \S5.

Let us also note a useful dual formulation of (\ref{Bpot}),
(\ref{basicpot}) which elucidates the monodromy effect introduced by
the wrapped brane.  Consider a D5-brane in type IIB string theory
wrapped on a two-cycle arising as the blowup cycle of a
supersymmetric $\mathbb{R}^3\times S^1/\mathbb{Z}_2$ orbifold; this
is equivalent to a fractional D3-brane at the orbifold singularity.
There is a T-dual, ``brane box", description of this configuration,
in which the fractional D3-brane becomes a D4-brane stretched
between two NS5-branes on a T-dual circle (see e.g. \cite{Andreas}).
Moving in the $b$ direction through multiple periods in closed
string moduli space in the original description corresponds to
moving one of the NS5-branes around the circle, dragging the
D4-brane around with it so as to introduce multiple wrappings.  This
T-dual description makes the linear potential manifest; see Fig. 2.

\begin{figure}[t]
\begin{center}
\includegraphics[width=10cm]{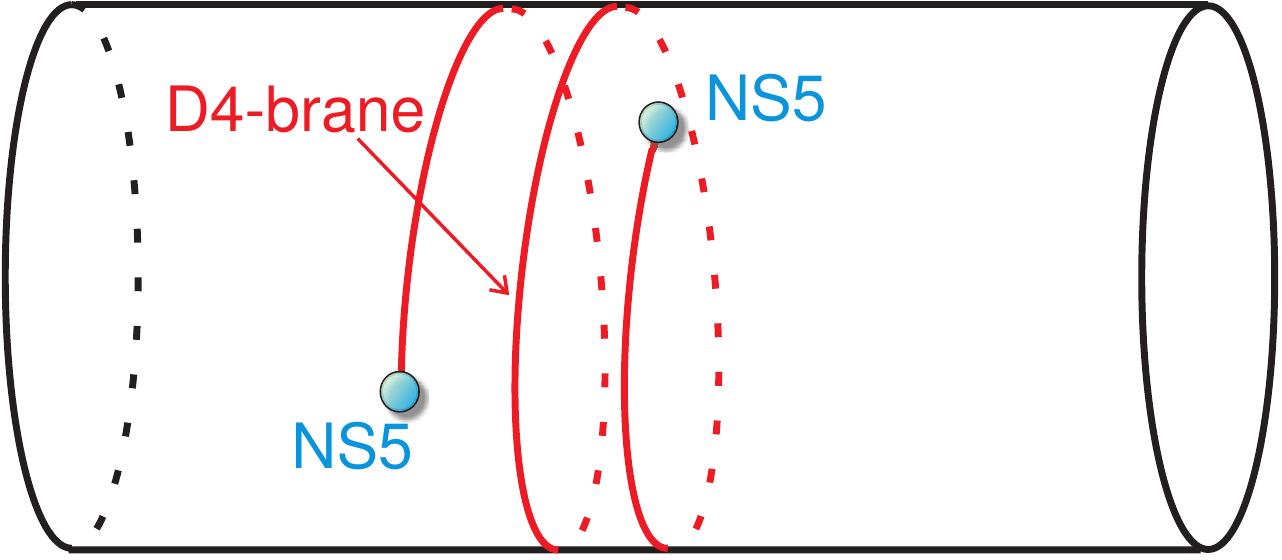}
\end{center}
\refstepcounter{figure}\label{winduptoy}

\vspace*{-.2cm} {\bf Figure~\ref{winduptoy}:} T-dual, ``brane box",
description of this configuration, in which the fractional D3-brane
becomes a D4-brane stretched between two NS5-branes on a T-dual
circle.  Moving in the $b$ direction through multiple periods in
closed string moduli space in the original description corresponds
to moving one of the NS5-branes around the circle, dragging the
D4-brane around with it so as to introduce multiple wrappings.
\end{figure}

\subsection{Basic Phenomenological Requirements}

Our candidate inflaton action takes the form
\be\label{basicinflac} S=\int d^4 x
\sqrt{g}\,\Bigl(\frac{1}{2}(\del\phi_a)^2-\mu_a^3\phi_a\Bigr)+{\rm
corrections} \ee
where we indicated corrections which we will analyze below,
suppressing them using symmetries, warping, and the natural
exponential suppression of nonperturbative effects.

In order to obtain 60 e-folds of accelerated expansion, inflation
must start at $\phi_a\sim 11 M_P$. In addition, the quantum
fluctuations of the inflaton must generate a level of scalar
curvature perturbation $\left.\Delta_{\cal R}\right|_{60}\simeq
5.4\times 10^{-5}$, with
\beq \left.\Delta_{\cal
R}\right|_{N_e}=\left.\sqrt{\frac{1}{12\pi^2}\,\frac{V^3}{M_{\rm
P}^6V'^2}}\right|_{N_e}
\eeq
This requires
\be\label{munumber}\mu_a\sim 6\times 10^{-4}M_P \ee
%


Given $f_a=\phi_a/a$ and the above results, the {\it number of
circuits} of the fundamental axion period $(2\pi)^2f_a$ required for
inflation is
\be\label{Nwgeneral} N_w=11 {M_P\over {f_a (2\pi)^2}} \ee
We will compute this number of circuits in each of the specific
models below. In the very simple case with all cycles of the same
size, this gives, using (\ref{samescalesMp}),
\be\label{Nwsamescales} N_w\sim 11\sqrt{6} {L^2\over{(2\pi)^2}}
~~~~{\rm (one ~ scale)} \ee
for $b$, while the requisite number of circuits for an RR inflaton
$c$ is larger by a factor $1/g_s$.

\subsection{Constraints on Corrections to the Slow-Roll Parameters}

Our next task is to ensure that the inflaton potential
$V_{inf}\approx \mu_a^3\phi_a$ is the primary term in the axion
potential.
All other contributions to the axion potential must make negligible
contributions to the slow roll parameters
\be\label{slowroll} \epsilon={M_{\rm P}^2\over 2}\left({V'\over
V}\right)^2 ~~~~~ \eta=M_{\rm P}^2 {V''\over V} \ee
A good figure of merit to keep in mind is that Planck-suppressed
dimension-six operators such as $V(\phi-\phi_*)^2/M_P^2$, with
$\phi_*$ a constant, contribute ${\cal O}(1)$ corrections to $\eta$.
In what follows, we will analyze the conditions for sufficiently
suppressing corrections to the slow-roll parameters.

Our specific setups discussed below will include reasonably generic
examples which naturally suppress these corrections well below the
one percent level, as is required in standard slow-roll inflation.
In other examples, instanton-induced sinusoidal corrections to the
potential lead to {\it oscillating} shifts in $\eta$ of order one.
Let us pause to assess the conditions on the slow-roll parameters in
monodromy-driven inflation.
In this class of models, the brane-induced inflaton potential is the
leading effect breaking the approximate shift symmetry in the
inflaton direction; other effects -- in particular, instantons, in
the case of our axion models -- produce {\it periodic} corrections
to the potential.  In general, such models can tolerate larger {\it
oscillating} contributions to $\eta$, as we now explain.


In the present situation, the corrections $\Delta\epsilon$ and
$\Delta\eta$ to the slow-roll parameters oscillate as a periodic
function of $a=\phi_a/f_a$ with period $(2\pi)^2$.  The potential
becomes steeper and flatter repeatedly during the evolution, and
because these two effects can compensate each other, it is worth
analyzing carefully what level of suppression of the amplitude of
$\Delta\eta$ is really necessary to ensure 60 e-folds of inflation
overall.

Let us simply give order-of-magnitude, parametric estimates for the
net effect of the steeper and flatter regions.  It would be
interesting to study this in more detail, with an eye toward
ancillary observational signatures which might arise in the power
spectrum of density perturbations.

The potential takes the form\footnote{Here for simplicity we neglect
terms proportional to $\phi_a {\rm cos}(\phi_a/2\pi f_a)$, as they
produce subdominant corrections to the slow roll parameters.}
\be\label{potoscillations} V=\mu_a^3\phi_a + \Lambda^4~{\rm
cos}\left({\phi_a\over {2\pi f_a}}\right) \ee
with $\Lambda$ a constant determined by the instanton action.

The second term yields an oscillating contribution to $\eta$ given,
for $\Lambda \ll \mu_a^3\phi_a$, by
\be\label{osceta} \eta=M_p^2 \left({1\over{2\pi
f_a}}\right)^2{\Lambda^4\over{\mu_a^3\phi_a}}{\rm cos}\left(
{\phi_a\over {2\pi f_a}}\right) \ee
The condition that the slope $V'(\phi_a)$ be non-negative can be
written as
\be\label{slopecond} 11 \eta {2\pi f_a\over M_P} \le 1 \ee
where we used (\ref{osceta}) and the fact that $\phi_a\le 11 M_P$
during the 60 e-folds of inflation in our linear potential.

Let us assume that averaging over the oscillations, the system
remains in its slow-roll regime, and check the conditions for this
to be self-consistent. The average field velocity is then
\be\label{phiv} \dot\phi_a \simeq -{\mu_a^3\over{3H}}, \ee
and the time $\Delta t$ during a period $\Delta\phi_a\sim
(2\pi)^2f_a$ is of order ${\Delta\phi_a\over\dot\phi_a}\sim
3(2\pi)^2f_aH/\mu_a^3$. Using this and the fact that $\eta$ is of
order $\ddot\phi_a/H\dot\phi_a$, we obtain the change
$\Delta\dot\phi_+$ in the field velocity during the (half-)period in
which the potential is relatively steep:
\be\label{vplus} \Delta\dot\phi_+\sim |\eta| (2\pi)^2f_a H \ee
Similarly, on the flat regions of the potential,
$\ddot\phi_a+3H\dot\phi_a\simeq 0$, and we obtain
\be\label{vminus} \Delta\dot\phi_-\sim -(2\pi)^2f_a H  \ee
Thus, we see that the kinetic energy does not build up over each
full period of oscillation between steeper and flatter potential
energy -- which ensures that potential-energy dominated inflation
proceeds -- as long as $|\eta|\lesssim 1$.  Again, many of the
specific examples realizing axion monodromy inflation described
below naturally yield much smaller corrections to $\eta$, but this
possibility of larger oscillations in other examples is an
intriguing new element worth investigating further in future work.


\section{Necessary Conditions for Controlled Inflation}

So far, we have a candidate for inflation along the direction
$\phi_a$, with potential $V_{inf}\approx \mu_a^3\phi_a$. We must now
ensure that the proposed inflaton action (\ref{basicinflac}) indeed
arises in a consistent and
controllable string compactification. This entails a series of
nontrivial conditions
dictated not directly by observations, but by our goal of producing
a consistent and {\it computable} string realization. We first
briefly summarize these requirements, then, in the following
subsections, show how each of them can be met. As in
\cite{Freese:1990rb}, we will use the natural exponential
suppression of instanton corrections to the axion potential.

The first, rather obvious condition is that the axion $a$ which is
to serve as the inflaton is actually part of the spectrum.  This
constrains the structure of the orientifold action used in moduli
stabilization; however, we expect that some suitable modes do
survive a generic orientifold projection. Next, we must demonstrate
that the proposed inflaton potential is in fact the dominant
contribution to the total potential for $a$: additional effects in
the compactification must make subleading contributions to the axion
potential.  Specifically, couplings to fluxes and periodic
contributions from instantons (worldsheet instantons and D-brane
instantons, in the cases of $b$ and $c$, respectively) must
therefore be controlled or eliminated. Next, we must show that the
energy stored in the axion does not source excessive distortion of
the local geometry near the wrapped branes. Finally, the inflaton
potential must remain subdominant to the moduli-stabilizing
potential, and shifts in the moduli during inflation must not give
large corrections to the inflaton potential.



\subsection{Axions and the Orientifold Projection}

We must first ensure that the axions $b,c$ of interest are part of
the spectrum.  That is, the orientifolds which are crucially used in
moduli stabilization (or their generalizations in F-theory) must
project in the required modes.  Some of the conditions for this in
the case of type IIB Calabi-Yau O3/O7 orientifolds appear in
\cite{GL,Grimm}, where the corresponding multiplets consist of $b$
and $c$ fields descending from K\"ahler moduli hypermultiplets in
the ``parent" unorientifolded Calabi-Yau manifold.

The worldsheet orientation reversal $\Omega$ which is part of every
orientifold projection acts with a (-1) on the Neveu-Schwarz
two-form potential $B_{MN}$.  However, orientifolds typically
include a geometric projection -- a reflection $I_{9-p}$ on some
$9-p$ directions -- at the same time.  Two simple situations in
which axions are projected in are the following.  First, a $B_{MN}$
field with one leg along the orientifold $p$-plane and the other
transverse to it will be projected in by the full $\Omega I_{9-p}$
action. Second, the orientifold may exchange two separate cycles
$\Sigma_1$ and $\Sigma_2$, independent in homology in the covering
space, into each other. This projects in one combination of the two
axions of the parent theory.

\subsection{Conditions on the Potential}

A generic string compactification will generate additional
contributions to the potential for $\phi_a$ going beyond the
candidate inflaton potential (\ref{basicpot}) (\ref{genpot}).  In
this subsection, we will describe the conditions for these
corrections to be consistent with inflation.


\subsubsection{Conditions on Flux Couplings}

We must first ensure that background fluxes do not couple to the
putative inflaton in such a way as to introduce problematic
contributions to the potential.  Ramond-Ramond fluxes $\tilde F_q$
include Chern-Simons corrections of the form $B_2\wedge F_{q-2}$ and
$C_{q-3}\wedge H_3$. These contributions, if present in the flux
compactification being used to stabilize the moduli, yield masses
for the corresponding components of $b$ and $c$ through the terms
proportional to $|\tilde F_q|^2$ in the ten-dimensional Lagrange density.

The extra contributions to the generalized field strengths give
contributions of the form
\be\label{fluxcouplingsB} \int d^{10} x {\sqrt{g}\over
{16(2\pi)^7\alpha'^4}}|B_2\wedge F_p |^2  \ee
or, in the $C^{(p)}$ case,
\be\label{fluxcouplingsC} \int d^{10} x {\sqrt{g}\over
{16(2\pi)^7\alpha'^4}}|C_p\wedge H_3 |^2 \ee
to the effective action (for definiteness we have given the
normalizations for the case of $|\tilde F_5|^2$ in type IIB).  It is
worth emphasizing that in type IIB flux compactifications on
Calabi-Yau orientifolds, the class of fluxes that are consistent
with the no-scale structure derived in \cite{GKP,GL}, namely
imaginary self-dual fluxes, do not contribute to the axion
potential: the axionic fields enjoy a no-scale cancellation of their
contribution to the flux-induced potential \cite{GL}.

In more general models we will have to ensure that we can make
analogous choices of fluxes to remove flux contributions to the
axion potential.  If the wedge products (\ref{fluxcouplingsB}),
(\ref{fluxcouplingsC}) are nonzero, and if the relevant flux $F_p$
or $H_3$ contributes leading moduli-stabilizing terms of order the
barriers in ${\cal U}_{mod}$, then the corresponding axion may be
obstructed from being the inflaton.  As an example, consider the
case of a product manifold. The coupling (\ref{fluxcouplingsB})
scales like
\be \int d^{10} x {\sqrt{g}\over {16(2\pi)^7 \alpha'^4 }}
|F_p|^2|b/L^2|^2 \approx \int d^{10} x \sqrt{g}{3|F_p|^2\over
{8(2\pi)^7 \alpha'^4 }} \frac{\phi_b^2}{M_P^2} \label{fluxBmass}\ee
while the contribution of the $F_p$ flux to the moduli potential
scales like
\be\label{Fpbarrier} \int d^{10} x {\sqrt{g}\over
{2(2\pi)^7\alpha'^4}} |F_p|^2\sim \int d^4x \sqrt{g_4}\ {\cal
U}_{mod} \ee
%
Thus, a super-Planckian excursion of the $\phi_b$ field would lead
to a contribution (\ref{fluxBmass}) which would overwhelm the
moduli-stabilizing barriers.\footnote{There are interesting ideas
for obtaining a large field range via large-N gauge theory
\cite{Juan}, which on the gravity side might involve warped-down
flux-induced monodromy. This may provide a way to use flux couplings
to introduce an inflationary axion potential consistent with moduli
stabilization,  but this question requires further analysis.}
Similar comments apply to curvature couplings and generalized
fluxes.



\subsubsection{Effects of Instantons}

The effective action for axions is corrected by instanton effects.
Worldsheet instantons depend periodically on $b$ type axions, while
Euclidean D-branes (D-brane instantons) introduce periodic
dependence on the $c$ type axions (and non-periodic, exponentially
damped dependence on $b/g_s$). Both types of instantons are
exponentially suppressed in the size of the cycle wrapped by the
Euclidean worldsheet or worldvolume.\footnote{One may also consider
nonperturbative effects arising in Euclidean quantum gravity, as
explored in \cite{KLLS}; these are exponentially suppressed in the
controlled regime of weak coupling and weak curvature.}

First, consider the kinetic terms in the effective action.  These
take the form
\be\label{kineticcorr} {1\over 2}\int d^4x\ \sqrt{g}f_a^2(\del
a)^2\Bigl(1+\epsilon_1 f_{per}(a)\Bigr) \ee
where $f_{per}(a)$ is a periodic function of $a\simeq a+(2\pi)^2$
normalized to have amplitude 1. This changes the canonically
normalized field to be
\be\label{canfldcorr} \phi_a=f_a\int^a da'\sqrt{1+ \epsilon_1
f_{per}(a')} \ee
Suppressing corrections to the slow roll parameters requires
sufficiently small $\epsilon_1$. In terms of the bare canonically
normalized field $\phi_{a}^{(0)}$, our periodic function varies on a
scale of order $(2\pi)^2f_a$: $f_{per}=f_{per}(\phi_{a}^{(0)}/f_a)$.
Thus for small $\epsilon_1$, the potential expanded about a local
minimum $\phi_*$ of $f_{per}$ is of the form
\be\label{Vcorrkin} V_{inf}(\phi_a)\simeq
\mu_a^3\phi_a\Biggl(1+\epsilon_1 {(\phi_a-\phi_*)^2\over{(2\pi
f_a)^2}}\Biggr)= \mu^3\phi_a\Biggl(1+\epsilon_1\left({M_P\over {2\pi
f_a}}\right)^2 {(\phi_a-\phi_*)^2\over{M_P^2}}\Biggr) \ee
Thus if
\be\label{sufficientkin} \epsilon_1 \lesssim  10^{-2}\left({2\pi
f_a\over M_P}\right)^{2} \approx \frac{1}{4\pi^2N_w^2} \ee
then the instanton corrections to the kinetic terms do not affect
inflation, since the slow roll parameters $\epsilon={M_{\rm
P}^2\over 2}({V'\over V})^2$ and $\eta=M_{\rm P}^2 {V''\over V}$
remain of order $10^{-2}$.

Next, let us consider instanton corrections arising directly in the
potential energy term in the effective action. These we can write as
(using similar notation to that above)
\be\label{Vcorrinst} V_{inf}(\phi_a)\sim
\mu_a^3\phi_a\Bigl(1+\epsilon_2 g_{per}(\phi_a/f_a)\Bigr)+\epsilon_3
{h_{per}(\phi_a/f_a)\over{\alpha'^2}} \ee
As before, let us assess sufficient conditions on $\epsilon_2$ and
$\epsilon_3$ to ensure that instanton corrections to the slow-roll
parameters are negligible.  From the first term in
(\ref{Vcorrinst}), we see that
\be\label{sufficientVI} \epsilon_2 \lesssim 10^{-2}\left({2\pi
f_a\over M_P}\right)^{2} \approx \frac{1}{4\pi^2N_w^2} \ee
From the second term, we find
\be\label{sufficientVII} \epsilon_3 \lesssim 10^{-2}\left({2\pi
f_a\over M_P}\right)^{2} \left(V_{inf}\alpha'^2\right) \approx
\frac{V_{inf}\alpha'^2}{4\pi^2N_w^2}  \ee
Note that the conditions we have imposed here may be relaxed, as
discussed in \S2.4, because of the oscillatory nature of the
corrections.  We will obtain negligibly small corrections to $\eta$
in a simple subset of our specific examples below, but it is worth
keeping in mind the possibility of a larger oscillating contribution
in other examples.

So far we have enumerated conditions on the amplitudes $\epsilon_i,
i=1,2,3$ of various instanton contributions to the effective action.
In order to implement these conditions, we need to relate the
$\epsilon_i$ to parameters of the stabilized string compactification
in a given model.  An {\it exponentially} small coefficient
$\epsilon_i$ arises automatically if the instanton wraps a cycle
larger than the string scale. For instantons wrapping small cycles,
$\epsilon_i$ may still be small if the kinetic term is protected by
local supersymmetry in the region near the cycle, or if the
instanton dynamics is warped down. We will consider several of these
cases in the specific models discussed below.

\subsection{Constraints from Backreaction on the Geometry}

We obtained the effective potential from our wrapped fivebrane using
standard results from ten-dimensional string theory. A basic
condition for control of our models is the absence of backreaction
of the brane on the ambient geometry, so that this ten-dimensional
analysis is valid to a good approximation. In particular, the core
size $r_{core}$ of our wrapped brane, including the effects of the
axion,
must be smaller than the smallest curvature radius $R_\perp$
transverse to it in the compactification.

A single D5-brane is pointlike at weak string coupling, and a single
NS5-brane is string-scale in size. However, in our regime of
interest the branes in effect carry $N_w\sim a/(2\pi)^2$ units of
D3-brane charge.  $N_w$ D3-branes produce a backreaction at a length
scale $r_{core}$ of order
\be\label{coresize} r_{core}^4\sim 4\pi\alpha'^2 g_s N_w \ee
Thus in order to avoid significant backreaction on our
compactification geometry, we require
\be\label{coresizecond} N_w \ll {R_\perp^4\over{4\pi g_s\alpha'^2}}
\ee
The one-scale expression for $N_w$ derived in \S2.3\ suggests that
this condition will be straightforward to satisfy, since the right
hand side of (\ref{coresizecond}) is $\propto R^4$, while the
expression (\ref{Nwsamescales}) scales like two powers of the
relevant length scale in the problem.  However, fitting GUT-scale
inflation into a stabilized compactification requires high
moduli-stabilizing potential barriers, which puts constraints on how
large the ambient compactification may be.  We will implement this
condition in the specific models to follow.

\subsection{Constraint from the Number of Light Species}

A related but slightly more subtle condition concerns new light
species that arise in our brane configuration at large $b$ or $c$.
The effectively large D3-brane charge $N_w$ introduces of order
$N_w^2$ light species.  It is important to check the contribution
this makes to the renormalized four-dimensional Planck mass.  In the
regime $g_sN_w>1$, the effect of the D3-brane charge is best
estimated using the gravity side of the (cutoff) AdS/CFT
correspondence, following Randall and Sundrum \cite{RS}.  As just
discussed, in the regime (\ref{coresizecond}), the size $r_{core}$
of the gravity solution for the D3-branes is smaller than the
ambient size $L\sqrt{\alpha'}$ of the compactification.  This leads
to a negligible contribution to $M_P^2$.

\subsection{Consistency with Moduli Stabilization}

A  further condition is that our inflaton potential, which depends
on the moduli as well as on $\phi_a$, not exceed the scale of the
potential barriers ${\cal U}_{mod}$ separating the system from
weak-coupling and large-volume runaway directions in moduli space:
\be\label{basicVU} V_{inf}(\phi_a) \ll {\cal U}_{mod} \ee
Since our large-field inflation model has a GUT-scale inflaton
potential, this requires high moduli-stabilizing potential barriers.

One must also ensure that the shifts in the moduli induced by the
inflaton potential do not appreciably change the shape of the inflaton
potential: in other words, the moduli-stabilizing potential must not
only have high barriers, it must also have adequate curvature at its
minimum.  A self-consistent way to analyze such shifts is to use an
adiabatic approximation, in which the moduli $\sigma$ adjust to sit
in instantaneous minima $\sigma_{*}(\phi)$ determined by the
inflaton VEV: \be\label{shift}
\partial_{\sigma}\Bigl(V_{inf}(\phi,\sigma)+{\cal U}_{mod}(\sigma)\Bigr)\Bigl|_{\sigma=\sigma_{*}(\phi)}=0 \ee One
then computes the correction this introduces in the inflaton
potential $V(\phi,\sigma_{*}(\phi))$  and checks whether this
correction is negligible.

For moduli stabilization mechanisms which use perturbative effects,
this condition is satisfied provided (\ref{basicVU}) holds, as
explained in \S2.4.2\ of \cite{Silverstein:2008sg}. Let us briefly
summarize this here. The volume and string coupling are exponentials
in the canonically normalized fields $\sigma$; for example the
volume is ${\cal V}\alpha'^3={\cal
V}_*e^{\sqrt{3}\sigma_v/M_P}\alpha'^3$ where ${\cal V}_*$ is the
stabilized value of the volume and $\sigma_v$ is the
canonically-normalized field describing volume fluctuations. In
perturbative stabilization mechanisms, the leading terms in the
moduli potential scale like powers of $L\equiv {\cal V}^{1/6}$:
schematically,
\be\label{powerstructure} {\cal U}_{mod}\sim \sum_n {c_n\over L^n}
\ee
where the coefficients $c_n$ depend on other moduli in a similar
way. Putting these two facts together, we see that derivatives of
the inflaton potential $V_{inf}$ and the moduli-stabilizing
potential ${\cal U}_{moduli}$ with respect to $\sigma/M_P$ scale
like the potential terms themselves.  Combining the tadpole from the
inflaton potential $V_{inf}$ with the mass squared from the moduli
potential ${\cal U}_{mod}$ yields the moduli shifts
\be\label{modshiftspower} {\sigma\over M_P} \sim {V_{inf}\over {\cal
U}_{mod}}  \ee
Plugging this back into the potential yields corrections which
change its shape.  However, these are small, giving corrections to
$\eta$ of order $\eta V_{inf}/{\cal U}_{mod}$.

For mechanisms we will study which employ exponential (e.g.
instanton) effects to stabilize the volume \cite{KKLT,LV}, the
structure of the potential is schematically \cite{GL,Grimm}
\be\label{expstructure} {\cal U}_{mod}\sim \sum_{n,m} {c_n\over
L^n}\,{\rm{exp}}\Bigl(-C_m \left(L^4/g_s+\tilde\gamma
b^2\right)\Bigr) \ee
where $\tilde\gamma$ is a factor of order unity.  From this we can
analyze two important consistency conditions.

First, let us discuss the moduli shifts.  In the expression
(\ref{expstructure}), there are still power-law prefactors in the
potential (arising from the rescaling of the potential to Einstein
frame) which lead to a similar suppression in the tadpoles for the
volume and the string coupling as in the perturbatively-stabilized
case. Furthermore, there are additional contributions to the masses
of the moduli from differentiating the exponential terms which can
enhance the masses relative to the perturbatively-stabilized case,
further suppressing the tadpoles.

Second, it will be important to keep track of which combinations of
geometrical moduli and axions are stabilized by a given
moduli-stabilization mechanism in calculating the slow-roll
parameters.   In the scenario \cite{KKLT}, a combination of the
volume and $b$ type axions of the form $L^4/g_s+\tilde\gamma b^2$ is
what is stabilized by ${\cal U}_{mod}$.  This leads to an $\eta$
problem for inflation along the direction of any Neveu-Schwarz axion
$b$, analogous to the $\eta$ problem identified in \cite{KKLMMT}. In
this class of models, we will therefore be led to consider instead
RR two-form axion inflation.

\section{Specific Models I: Warped IIB Calabi-Yau Compactifications}

In this section and in \S5, we implement our basic strategy in
several reasonably concrete models, imposing the consistency
conditions delineated above.  This is an important exercise,
necessary in order to ensure that it is indeed possible to satisfy
all the conditions together.  Needless to say, there are many ways
to generalize -- and potentially simplify -- these constructions,
and we will indicate along the way some further directions for model
building.

There are two classes of examples which differ in which combinations
of scalar fields are stabilized by the moduli-fixing potential.  In
the case of mechanisms such as those outlined in
\cite{GKP,KKLT,LV,KL}\ which employ nonperturbative effects in a
low-energy supersymmetric formulation, the moduli potential
stabilizes a combination of the geometric and axion modes.  In
perturbative stabilization mechanisms such as those outlined in
\cite{MSS,Riemann,DGKT,Silverstein:2007ac}, the volume and other
geometrical moduli are directly stabilized. These latter cases will
be discussed in \S5.


One canonical class of examples arises in warped flux
compactifications of type IIB string theory on orientifolds of
Calabi-Yau threefolds.  After a telegraphic review of the resulting
low-energy supergravity, we show that nonperturbative stabilization
of the K\"ahler moduli leads to  an $\eta$ problem for a candidate
inflaton $b$ descending from $B^{(2)}$.  We then demonstrate that
this problem is absent when $C^{(2)}$ is the inflaton, and
furthermore show that the leading remaining dependence of the
potential on $c$, from Euclidean D1-branes, may be naturally
exponentially suppressed. Inflation driven by a wrapped NS5-brane
which introduces monodromy in the RR two-form axion direction is
therefore a reasonably robust and natural occurrence in warped IIB
compactifications.

\subsection{Multiplet Structure, Orientifolds, and Fluxes}

Consider a compactification of type IIB string theory on a
Calabi-Yau threefold.  The resulting four-dimensional ${\cal N}=2$
supergravity contains $h^{1,1}+1$ hypermultiplets, one of which is
the universal hypermultiplet containing the axio-dilaton $\tau$.
The remaining hypermultiplets have as bosonic components
$b^A,c^A,{\rm Re}\,T^A,{\rm Im}\,T^A\equiv \theta^A$, where
$\theta^A=\int_{\Sigma_A^{(4)}}C_4$, and $T^A$
is the ${\cal N}=1$ complexified K\"ahler modulus, defined more
carefully below.  The axions suitable for monodromy inflation with
wrapped fivebranes are in the $(b^A,c^A)$ half of these
hypermultiplets.  The overall volume and other size moduli are
contained in the $T^A$.

We now consider orientifold actions, which break ${\cal N}=2 \to
{\cal N}=1$ and play an important role in moduli stabilization. We
will particularly focus on orientifold actions whose fixed loci give
O3-planes and O7-planes. For the orientifold action we take
\be\label{Oaction} {\cal O}= (-1)^{F_L} \Omega \sigma \ee where
$\sigma$ is a holomorphic involution of the Calabi-Yau, under whose
action the cohomology groups split as: \be\label{split}
H^{(r,s)}=H^{(r,s)}_+\oplus H^{(r,s)}_- \ee We correspondingly
divide the basis $\omega_A, A=1,\ldots h^{1,1}$ into $\omega_\alpha,
\alpha=1,\ldots h^{1,1}_+ $ and $\omega_I, I=1,\ldots h^{1,1}_-$. As
explained in detail in \cite{GL}, half of the fields are invariant
under the orientifold projection and are kept in the
four-dimensional theory. Specifically, K\"ahler moduli $T^\alpha $
corresponding to even cycles and axionic moduli $G^I = c^I - \tau
b^I$ corresponding to odd cycles survive the projection.

Let us indicate two classes of odd cycles we can project in by
orientifolding Calabi-Yau manifolds. The first, considered in
\cite{GL,Grimm}, consists of zero-size cycles which intersect the
orientifold fixed plane in a locus of real dimension one.  In this
case, the orientifold can project in $B_{MN}$ and $C_{MN}$ with
their legs oriented so that $M$ (say) is parallel and $N$ transverse
to the orientifold fixed plane. The size modulus for the two-cycle
is projected out in this case. The second construction arises when
the orientifold maps two separate cycles $\Sigma_1$ and $\Sigma_2$,
independent in homology in the covering space, into each other. In
this situation, the size modulus $T_+$ of the even combination
$\Sigma_+$ of the two two-cycles is projected in, while the $G$
modulus $G_-$ of the odd combination $\Sigma_-$ of the two-cycles is
projected in.  It is important to note that this requires the sizes
$v_1$ and $v_2$ of the two-cycles in the covering Calabi-Yau space
to be the same: $v_1=v_2\equiv {1\over 2}v_+$. The odd volume
modulus $v_-=v_1-v_2$ -- the difference in size of the two-cycles --
is projected out. Note that this allows for a situation in which
there are no small geometrical sizes anywhere in the orientifold, if
$v_+$ is large.  In particular, as long as $v_+$ is large, the fact
that $v_-$ is zero does not indicate the presence of any small
curvature radii or small geometrical sizes in the compactification.

\begin{figure}[!ht]
\begin{center}
\includegraphics[width=10cm]{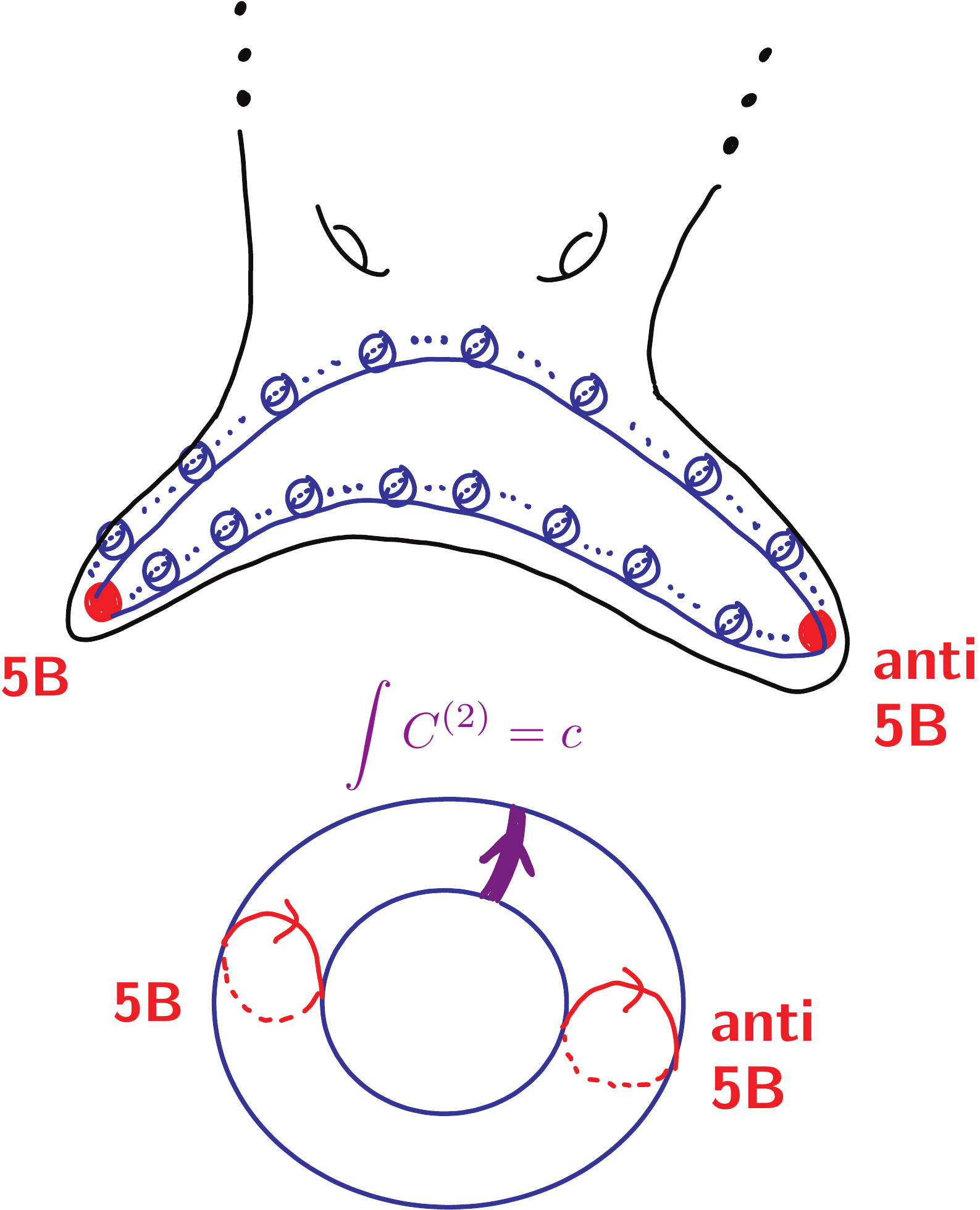}
\end{center}
\refstepcounter{figure}\label{snake}

{\bf Figure~\ref{snake}: }{\textbf{Schematic of
tadpole cancellation.}} Blue: Two-real-parameter family of
two-cycles $\Sigma_1$, drawn as spheres, extending into warped
regions of the Calabi-Yau. Red: We have placed a fivebrane in a
local minimum of the warp factor, and an anti-fivebrane at a distant
local minimum of the warp factor.  In the lower figure, $\Sigma_1$ is drawn as the cycle threaded by $C^{(2)}$, and global tadpole cancellation is manifest.
\end{figure}

Let us consider the latter `free exchange' case for definiteness.
In order to straightforwardly satisfy Gauss' law in the
compactification, it is simplest to consider two families of
two-cycles $\Sigma_1$ and $\Sigma_2$, extending into warped regions
of the parent Calabi-Yau. Within each family, place a fivebrane in a
local minimum of the warp factor, and an anti-fivebrane at a distant
local minimum of the warp factor.  The orientifold exchanges the two
families, yielding families of (anti)invariant two-cycles $\Sigma_+ (\Sigma_-)$.
The warped fivebrane, with its monodromy in the axion direction,
provides our candidate inflationary potential energy. This is
illustrated schematically in Fig.~\ref{snake}.



As a standard example, we may consider a warped throat which is
approximately given by $AdS_5 \times X_5$, where $X_5$ is an
Einstein space, the two factors have common curvature radius $R\sim
L\sqrt{\alpha'}$, and the throat is cut off in the IR and UV
\cite{RS,KS,GKP}.
\be\label{warpedmet}ds^2 = e^{2A(r)}\eta_{\mu\nu}dx^\mu dx^\nu +
e^{-2A(r)}\left(dr^2+r^2 ds^{2}_{X_5}\right) \ee
with warp factor $e^{A(r)}\sim r/R$.

To complete the definition of the K\"ahler moduli, we first define
the K\"ahler form $J=v_{\alpha}\omega^{\alpha}$. The
compactification volume ${\cal V}\alpha'^3$ satisfies\footnote{This
formula follows our convention (\ref{alphapr}); another common
convention is to define the volume in units of
$l_s=2\pi\sqrt{\alpha'}$ (see the appendix of \cite{FixingAll}).}
\be\label{volume} {\cal
V}=\frac{(2\pi)^6}{6}c^{\alpha\beta\gamma}v_{\alpha}v_{\beta}v_{\gamma}
\ee
where $c^{\alpha\beta\gamma}$  are the triple intersection numbers.
Then the complexified K\"ahler modulus is given by~\cite{GL}
\be\label{kah}
T_{\alpha}=\frac{3}{4}c^{\alpha\beta\gamma}v_{\beta}v_{\gamma}+\frac{3}{2}i\theta_{\alpha}+\frac{3}{8}e^{\phi}c^{\alpha
I J}G_I (G-\bar{G})_J \ee
In the instructive simple case where $h^{(1,1)}_+=1$ (so that the
index $\alpha$ takes a single value, $L$, corresponding to the
overall volume modulus), the classical K\"ahler potential for the
sector descending from non-universal hypermultiplets takes the form
\be\label{kp} {\cal K}=-3~ {\rm log}\left( T_L+\bar{T}_L
+\frac{3}{2}e^{-\phi}c^{LIJ}b_I b_J\right) +\ldots \ee
The quantity inside the logarithm depends only on the overall volume
and string coupling.

Moduli stabilization is essential for any realization of inflation
in string theory, and we must check its compatibility with inflation
in each class of examples. In type IIB compactifications on
Calabi-Yau threefolds, inclusion of generic three-form fluxes
stabilizes the complex structure moduli and dilaton \cite{GKP}.  A
subset of these three-form fluxes -- imaginary self-dual fluxes --
respect a no scale structure \cite{GKP,Grimm}. This suffices to
cancel the otherwise dangerous flux couplings described in \S3.2.1.

\subsection{An Eta Problem for B}

In this class of compactifications, however, the stabilization of
the K\"ahler moduli leads to an $\eta$ problem in the $b$ direction.
This problem arises because the nonperturbative effects (e.g. from
Euclidean D3-branes or strong dynamics on wrapped sevenbranes)
stabilize the K\"ahler moduli \cite{KKLT}\ $T^\alpha$ rather than
directly stabilizing the overall volume ${\cal V}\alpha'^3$.


Consider a setup with one or more D5-branes wrapping a curve
$\Sigma_I^{(2)}$; as already explained, $b_I$ is the candidate
inflaton in this case.
Now, the action for a Euclidean D3-brane wrapping the even
four-cycle $\Sigma_L^{(4)}$ is proportional to $T_L$, so the
nonperturbative superpotential depends specifically on the K\"ahler
modulus $T_L$,\footnote{We will soon consider the possibility of
axion dependence in the prefactor $A$; $a_L$ is a constant.}
\be\label{ed3} W_{ED3}=Ae^{-a_L T_L} \ee
On the other hand, the compactification volume involves a
combination (\ref{kp})  of the K\"ahler modulus and the would-be
inflaton $b_I$. The volume appears in the four-dimensional
potential, as usual, through the rescaling to Einstein frame;
equivalently, the volume ${\cal V}$ appears in the F-term potential
via the prefactor $e^K$. This inflaton-volume mixing is exactly
analogous to the problem encountered for D3-brane inflatons in
\cite{KKLMMT}; just as in that case, expansion of the potential
around the stabilized value of $T_L$ immediately reveals a
Hubble-scale mass for the canonically-normalized field $\phi_b$
corresponding to the axion $b$.  Hence $\eta \sim 1$, preventing
prolonged inflation.\footnote{Note that this is {\it not} an
oscillating contribution to $\eta$, and hence must be suppressed
well below ${\cal O}(1)$.}

We remark that this problem is apparently absent for the case of a
perturbatively-stabilized volume.  Moreover, because the volume
depends on $b^I$ but not on $c^I$, inflaton-volume mixing is also
not a problem for a model in which $c^I$ is the inflaton, which we
now consider in detail.\footnote{In \cite{LustII} it was recognized
that $b^I$ receives a mass from the leading nonperturbative
stabilization effects, whereas $c^I$ does not.}

\subsection{Instantons and the Effective Action for RR axions}

We are now led to consider a compactification on an orientifold of a
Calabi-Yau, in which one or more NS5-branes wrap a curve
$\Sigma_I^{(2)}$, and the leading moduli-stabilizing effects from
fluxes and Euclidean D3-branes -- or gaugino condensation effects --
do not contribute to the potential for $c^I$.  The next task is to
determine whether there are any further contributions to the
inflaton potential which might lead to overly strong dependence on
our candidate inflaton direction $c^I$.
In particular, Euclidean D1-branes, when present, introduce
sinusoidal contributions.  So we must study and control the effects
of Euclidean D1-branes.

\subsubsection{Instanton Contributions to the Superpotential}

A priori, one might expect the superpotential to take the schematic
form
\be\label{wis} W=\int (F_3-\tau H_3)\wedge\Omega+Ae^{-a_L T_L}+Be^{-
\tilde a( v_+-G_-/(2\pi)^2)}+Ce^{- a_L T_L-a_L G_-/(2\pi)^2} +\Delta
W(T_+),\ee
where for simplicity of presentation we have restricted attention to
a single pair of cycles freely exchanged by the orientifold, with
corresponding fields $T_+, G_-$, as well as an additional four-cycle
$\Sigma_L^{(4)}$ associated with the overall volume; the prefactors $A, B, C$ are constants.

Let us discuss each term in turn.  The first two terms represent the
moduli-stabilizing contributions of \cite{GKP}\ and \cite{KKLT}; we
will discuss additional features arising in the case of the large
volume scenario \cite{LV}\ below. The next putative term represents
the contribution of Euclidean D1-branes. Here $v_+$ is the volume of
the orientifold-even two-cycle $\Sigma^{(2)}_{+}$; as explained in
\cite{GL}, $v_+$ belongs to a linear multiplet, not a chiral
multiplet.  Holomorphy therefore forbids the superpotential from
depending on $v_+$ (said another way, the proper K\"ahler
coordinates are $T_L, T_+$, which are four-cycle volumes), but at
the same time any Euclidean D1-brane effect must vanish at large
volume. So the final term in (\ref{wis}) must be absent
\cite{Witten}.

The next term, proportional to $C$, represents Euclidean D1-brane
corrections to the Euclidean D3-brane action (which we will refer to
as ED3-ED1 contributions), in the case without wrapped sevenbranes
on the corresponding four-cycle. (We will discuss the case of strong
dynamics on sevenbranes further below.)   When present, this arises
from a Euclidean D1-brane dissolved as flux in a Euclidean D3-brane;
see e.g. \cite{MMMS}.  Such a contribution requires that (a
supersymmetric representative of) the two-cycle carrying our
$C^{(2)}$ axion be embedded in the (supersymmetric) four-cycle
wrapped by the original Euclidean D3-brane.  When the cycles are
configured in this way, the resulting dependence of the
superpotential on $c^I$ appears to be unsuppressed compared to the
leading moduli-dependence, and in those cases one should worry that
the moduli-stabilizing superpotential gives the inflaton a large
mass-squared, of order ${\cal U}_{mod}/(2\pi f_a)^2> H^2$.

We could attempt to control this effect using warping. That is, if
the two-cycle in question, and all two-cycles in its homology class,
are localized in a warped region, then the coefficient $C$ in
(\ref{wis}) is suppressed, on dimensional grounds, by three powers
of the warp factor $e^{A_{top}}$ at the top of the two-cycle fixed
locus in the throat
\be\label{Bwarp} C\sim e^{3A_{top}}\quad. \ee
Since this contribution was marginally dangerous to begin with, a
modest warp factor satisfying \be\label{Bwarpcond}
e^{3A_{top}}<\Delta\eta(2\pi f_a/M_P)^2, \ee with $\Delta\eta$
constrained as described in \S2.4, suffices to avoid significant
contributions to the slow-roll parameters.

However, examples generating this contribution to the superpotential
are tightly constrained with respect to the basic backreaction
condition (\ref{coresizecond}), as follows.
A computation of the kinetic term for $c$ shows\footnote{This
analysis proceeds as in \cite{RS}, with $c$ a bulk scalar field in
the throat, or alternatively by the method reviewed in \S2.1.} that
the axion decay constant $f_c$ is proportional to the maximal warp
factor arising in the homology class of the corresponding two-cycle:
$f_c\sim e^{A_{top}}\hat f_c$.  This implies $N_w\sim 11
e^{-A_{top}}/(\hat f_c (2\pi)^2)$. Putting this together with
(\ref{coresizecond}), we obtain the constraint $e^{A_{top}}>11\hat
f_c/(\pi g_s M_P)$.  But the condition (\ref{Bwarpcond}) is
equivalent to the condition $e^{A_{top}}<(2\pi)^2\Delta\eta \hat
f_c^2/M_P^2$. Together these would require $\Delta\eta>11
M_P/(4\pi^3 g_s^2\hat f_c)$.


Because of these issues, we will consider examples where the
dangerous ED3-ED1 terms do not arise.  One situation in which this
occurs naturally is the following.  Consider, as in \cite{KKLT}, the
case that the moduli-stabilizing nonperturbative superpotential
arises from gaugino condensation on sevenbranes wrapping
four-cycles.  In that situation, the physics below the KK scale of
the four-cycles is given by pure ${\cal N}=1$ $SU(N_L)$
supersymmetric Yang-Mills theory.   In terms of its holomorphic
gauge coupling $\tau_{YM}=\theta/(4\pi)+i/g_{YM}^2$, this theory has
an exact superpotential of the form \cite{Intriligator:1995au}\
\be\label{gaugecondW} W=\Lambda^3 = A
e^{{8\pi^2\over{N_L}}i\tau_{YM}} \ee
%
In order to determine the dependence of our superpotential on $T_L$
and $G_-$ (and in general on other moduli), we must determine the
Yang-Mills gauge coupling, including all significant threshold
corrections to it at the KK scale.  The Yang-Mills gauge coupling on
the sevenbranes is classically given by $8\pi^2/g_{YM}^2= 2\pi\,
{\rm Re}\,T_L$ (in the absence of magnetic two-form flux on the
D7-branes~\cite{LustD7with2formflux}), and similarly for other
four-cycles in cases with more K\"ahler moduli.  Comparing to
(\ref{gaugecondW}), we can identify the parameter $a_L$ in
(\ref{wis}) as $2\pi/N_L$ for this case.

The holomorphic gauge coupling function $\tau_{YM}$, like the
superpotential itself \cite{Witten}, is constrained by holomorphy
combined with the condition that in our weakly-coupled regime, all
nonperturbative corrections decay exponentially as the curvature
radii grow.  This means that $\tau_{YM}$, like $W$, cannot develop
pure ED1 corrections of order $e^{-2\pi G_-/N_L}$; instead, the
leading correction to the gauge coupling function must be
exponentially suppressed in $T_L$.  Plugging this into
(\ref{gaugecondW}), we see that the leading corrections from such
threshold effects to the superpotential itself are exponentially
suppressed relative to the leading moduli-stabilizing terms in
(\ref{wis}).  In particular, the coefficient $C$ in (\ref{wis}) is
negligible in this setup.

\subsubsection{Instanton Contributions to the K\"ahler potential}

Next, we note that the corrected K\"ahler potential can be written
schematically as follows\footnote{See e.g. \cite{Camara} for related
work in the type I string.} (with similar terms depending on $T_+$):
\be\label{kprob} {\cal K}=-3 ~{\rm log}\left( T_L+\bar{T}_L
+\frac{3}{2}e^{-\phi}c^{LIJ}b_I b_J+C_+\,{\rm Re}\, e^{-2\pi
v^+-G_-/(2\pi)}\right) +\ldots \ee
In contrast to the holomorphic gauge coupling and superpotential
just discussed, the K\"ahler potential is not protected by
holomorphy. The dependence on $\phi_c$ arising through the appearance
of $G_-$ in (\ref{kprob}) can naturally be suppressed to the
necessary extent by using the exponential suppression in the size
$v_+$ of the two-cycle. The ED1 contribution here yields a shift of
$\eta$ of order
\be\label{Kahlereta} \Delta\eta \sim {{\cal
U}_{mod}\over{V_{inflation}}}(2\pi)^2 {C_+\over g_s} e^{-2\pi v_+}
\ee
This is straightforward to suppress with a modest blowup $v_+$ of
the even cycle.  (Alternatively, one may consider the possibility
discussed in \S2\ of larger oscillating contributions to $\eta$.)

In general, there are several mechanisms one can consider for
suppressing instanton effects, including use of local symmetries,
warping, and (as just mentioned) exponential suppression of
instanton effects with the geometrical sizes of cycles they wrap.
Let us elaborate on the latter approach, which is likely to be the
generic situation.

In the KKLT mechanism of moduli stabilization, nonperturbative
effects are used to stabilize K\"ahler moduli. Consider using this
mechanism to stabilize the K\"ahler modulus $T_+$ corresponding to
the geometric size of the two-cycle wrapped by our NS5-brane
(strictly speaking, the K\"ahler modulus corresponds to the size of
the dual four-cycle, and we implicitly use the relation between the
$T_\alpha$'s and $v_\alpha$'s.)  In doing this, we must keep both
$T_+$ and the overall volume sufficiently small that the barrier
heights exceed the GUT scale of our inflaton potential, which takes
the form (\ref{basicpot}): \be\label{blowup} V(c) =
{\epsilon\over{g_s^2(2\pi)^5\alpha'^{(6-p)/2}}}\sqrt{v_+^{2}+c^2
g_s^2} \ee The $v_+$ dependence in (\ref{blowup})  tends to compress
the wrapped two-cycle, so a crucial consistency condition, as
discussed in \S3.2\ for the overall volume and dilaton, is that the
modulus $T_+$ be stabilized strongly enough so as not to shift in
such a way as to destabilize inflation.  As in \S3.2, we must
therefore compare the tadpole from (\ref{blowup}) with the scale of
the mass introduced by the moduli stabilization potential ${\cal
U}_{mod}$. Unlike the overall volume, $T_+$ need not be
exponentially related to its canonical field $\sigma_+$ if it makes
a subleading contribution to the physical volume ${\cal V}$
appearing in the K\"ahler potential, so we must assess its shift
separately.


It is convenient -- and equivalent -- to work out the shift $\delta
v_+$ of $v_+$ rather than that of the canonically normalized field
$\sigma_+$, and then substitute the result back into the potential
to determine the size of the resulting corrections to slow roll
parameters in the $\phi_c$ direction. We obtain from the added
exponential terms in ${\cal U}_{mod}$ the leading contributions to
the mass term for $\delta v_+$
\be\label{Ttwomass} \del^2_{v_+}{\cal U}_{mod}\sim \Bigl({\partial
T_+\over{\partial v_+}}\Bigr)^2{\cal U}_{mod}\sim
(c^{++L}v_L+2c^{+++}v_+)^2{\cal U}_{mod} \ee
The tadpole introduced by the expanding the inflaton potential
(\ref{blowup}) in powers of $v_+^2/(c g_s)^2$ is of order
\be\label{Ttwotad} \del_{v_+} V \sim V{v_+\over {(cg_s)^2}}  \ee
This leads to a shift in $v_+$ of order
\be\label{shiftv} \delta v_+\sim {V\over{{\cal U}_{mod}}}
{(v_+/c^2g_s^2)\over{(c^{++L}v_L+2c^{+++}v_+)^2}} \ee
and a corresponding correction to the full scalar potential of order
\be\label{shiftU} \Delta {\cal U}_{tot}\sim V {V\over{{\cal
U}_{mod}}}{(v_+/c^2g_s^2)^2\over{(c^{++L}v_L+2c^{+++}v_+)^4}} \ee
This shift is negligible since $1\ll v_+ \ll cg_s$.


Once the cycle $v_+$ is stabilized at a value larger than string
scale, the  Euclidean D1-brane corrections to the K\"ahler potential
are exponentially suppressed.  This provides a natural mechanism for
ensuring the conditions (\ref{sufficientkin}), (\ref{sufficientVI}),
(\ref{sufficientVII}) that $\epsilon_1,\epsilon_2$, and $\epsilon_3$
are sufficiently small.


\subsubsection{Effects of Enhanced Local Supersymmetry}

In some cases, the instanton corrections might be small without
blowing up the cycle $\Sigma_I^{(2)}$. There is ongoing research on
stringy instanton effects; a systematic understanding of these
effects would substantially improve our ability to build concrete
axion inflation models. In particular, in some recent works, desired
D-instanton corrections were difficult to obtain because of
cancellations arising from extra fermion zero modes
\cite{StringyInstantons}. For our purposes this cancellation is
advantageous; one situation in which it is particularly likely is
when the region near the two-cycle locally preserves extra
supersymmetry. A more specific setup of this sort is one in which
our two-cycle is locally a $\mathbb{C}^2/\mathbb{Z}_2$ orbifold
blowup cycle in a warped throat.   In particular, consider such an
orbifold singularity, with the fixed point locus extending up the
radial direction of the warped throat to a maximal warp factor
$e^{A_{top}}$ (see Fig.~\ref{snake}) and along a circle of size
$2\pi R$ in the internal $X_5$ directions.

In this case, in the local six-dimensional system the modulus $v_I$
corresponds to a geometrical blowup of the two-cycle, and is
linearly related to the canonically normalized scalar field (as can
be seen for example by its T-dual relation to relative positions of
NS5-branes).


In the case of a Klebanov-Strassler throat, for example, we can
orbifold to obtain a fixed point locus which extends radially up the
warped throat and along an $S^1$ within the internal $T^{1,1}$, as
follows. In the standard presentation of the deformed conifold,
\be\label{conifold} \sum_{i=1}^4 z_i^2=\varepsilon^2 \ee
we obtain this with an orbifold action under which
$(z_1,z_2,z_3,z_4)\to (-z_1,-z_2,z_3,z_4)$.  This system has ${\cal
N}=4$ supersymmetry locally, and the extended nature of the fixed
point locus of the orbifold implies the presence of bosonic and
fermionic zero modes corresponding to the collective coordinates
describing the instanton's position in the radial direction and
along the $S^1$ within the $T^{1,1}$.   For this configuration, we
note that using (\ref{fcis}), the axion decay constant is given by
%
\be\label{warpedsnake} \phi_c \sim {c\over\sqrt{\alpha'}}
e^{A_{top}}\left({R\over\sqrt{\alpha'}}\right)\sim M_P
e^{A_{top}}{cg_s\over L^2}\ee
where again we keep track of the maximal warp factor in the region
explored by the entire family of homologous blowup two-cycles, a
quantity which is determined by the way in which the warped throat
is connected to the rest of the compactification.  It is worth
emphasizing that the simplest methods we outlined above for
suppressing corrections to the slow-roll parameters do not require
warping of the entire family of two-cycles; one may simply take
$e^{A_{top}}\sim 1$ provided that the cycle wrapped by the NS5-brane
is stabilized at finite volume and that the moduli-stabilizing
nonperturbative effects arise from sevenbranes.

\subsection{Backreaction Condition}

Let us next address the question of backreaction of our wrapped
brane inside the warped Calabi-Yau.  The basic condition
(\ref{coresizecond}) becomes
\be\label{coresizeCY} N_w \ll \frac{\pi^3}{4} {\rm Re}(T_L) \ee
where we used the relation ${\rm Re}(T_L)={{\rm Vol}_4\over{(2\pi)^4
g_s}}$ between the chiral field $T_L$ and the size ${\rm
Vol}_4\alpha'^2$ of the corresponding four-cycle in the Calabi-Yau
(which we then identified with $L^4\sim(2R_\perp)^4/\alpha'^2$). Now
let us combine this with the condition that the moduli potential
barriers exceed the scale of our inflation potential. The scale of
the moduli-stabilizing barriers is given in terms of the rank of the
gauge group $N_L$ on the sevenbranes as
\be\label{rank} {\cal U}_{mod}\simeq {|A|^2\over {T_L^3 M_P^2}}
e^{-4\pi T_L/N_L}. \ee
Setting ${\cal U}_{mod}\ge V_{inf} \simeq 2.4\times 10^{-9} M_P^4$
yields the constraint
\be\label{TLmin} T_L \le   - {N_L\over{4\pi}}~{\rm
log}\left(2.4\times 10^{-9} T_L^3{M_P^6\over {|A|^2}}\right)
~~~~\Rightarrow ~~~~ N_w\ll -\frac{\pi^2}{16} N_L~{\rm
log}\left(2.4\times 10^{-9} T_L^3{M_P^6\over {|A|^2}}\right) \ee
%
%
%
%
%
%
%

The coefficient $A$ in the superpotential may depend holomorphically on complex structure moduli.  As in much of
the previous literature on KKLT moduli stabilization, we will take $A\approx M_P^3$. However, we should note a
standard subtlety with loop corrections to the gauge coupling and how it affects our considerations.  In our
system, the four-dimensional ${\cal N}=1$ supersymmetric Yang-Mills theory on the sevenbranes crosses over at
the KK scale $M_{KK}\sim (2\pi)g_s M_P/L^4$ to an eight-dimensional maximally supersymmetric gauge theory.  The
effective cutoff scale in the quantum field theoretic analysis of the ${\cal N}=1$ supersymmetric Yang-Mills
theory \cite{Intriligator:1995au,SV}\ is therefore $M_{KK}$. This might naively suggest $A\sim M_{KK}^3$, but
this would not be holomorphic. The holomorphy of the superpotential (\ref{gaugecondW}) may be maintained by an
appropriate redefinition of the chiral superfield $T_\alpha$ appearing in the gauge coupling function for the
sevenbrane stack wrapping the four-cycle $\Sigma^{(4)}_\alpha$. This in turn introduces a shift proportional to
$-{N_\alpha\over{2\pi}}{\rm log}(T_\alpha+\bar T_\alpha)$ (plus a constant) in the argument of the logarithm
appearing in the K\"ahler potential (\ref{kp}).  Since we require relatively high moduli-stabilizing barriers,
our system lies close to the interior of K\"ahler moduli space, and this shift can become significant depending
on the details of the example.  A preliminary investigation suggests that with reasonable numbers the effect is
(barely) neglibigle in the stabilization of $T_\alpha$, and that more generally it may in fact push $T_\alpha$
to larger values at fixed, high, barrier heights.  Overall, we find that with modest choices of $N_L$, our
system can tolerate hundreds of circuits of the basic axion period.

\subsubsection{The Large Volume Scenario}

In the large volume mechanism \cite{LV}\ for stabilizing Calabi-Yau flux compactifications, both power-law and
nonperturbative terms play a role in stabilizing the K\"ahler moduli.  In this setting, one can increase the
volume, maintaining the required high barrier heights, by increasing $W_0$.  This provides another method for
ensuring satisfaction of the backreaction constraint.

\subsection{Numerical Toy Examples}

Our analysis of the conditions for inflation suggests that they are
reasonably straightforward to satisfy. Because the full potential is
somewhat complicated, it is worth checking numerically how the
scales work out in a four-dimensional supersymmetric effective
action which encodes the essence of our mechanism, including the
basic structures required for moduli stabilization and inflation. We
will therefore consider the four-dimensional action descending from
a compactification with the minimal possible content --
$h_+^{1,1}=2$ and $h_-^{1,1}=1$ -- required for the mechanism
described above, including the effects of the orientifold action.


%
We will consider setups of the sort described in the previous
subsection, with two size moduli denoted by an index $\alpha=L$ for
the overall volume mode and $\alpha=+$ for the even combination of
cycles under an orientifold action.  (The odd combination of cycles,
as described above, supports our $C^{(2)}$ inflaton field, $c_-={\rm
Re}[G_-]$.)



To mock this up, guided by the structure of orientifolds of
Calabi-Yau manifolds such as $\mathbb{P}^4_{11169}$
and $\mathbb{P}^4_{13335}$, we define a class of toy models by a
classical K\"ahler potential of the form
\be K=-2\ln{\cal V}_E=- 2\ln\left\{(T_L+\bar
T_L)^{3/2}-\left[T_++\bar T_++\frac{3}{8}g_s c_{+--}\,(G_-+\bar
G_-)^2\right]^{3/2}\right\}\quad. \ee
%
plus contributions depending on the dilaton and complex structure
moduli, where we defined
\be {\cal V}_E=\frac{L^6}{g_s^{3/2}(2\pi)^6}=\frac{\cal
V}{g_s^{3/2}(2\pi)^6}\quad. \ee
From the requirement of getting a positive-definite kinetic term for
$G_-$ we deduce $c_{+--}>0$ and for the following examples we choose
for convenience $c_{+--}=+1$.  We also include corrections to $K$
of the form given in (\ref{kprob}).

The superpotential we take to be of a generalized KKLT-type
structure \cite{KKLT,KL}\
\be\label{WKKLT} W=W_0+A_+e^{-a_+ T_+}+\left\{\begin{array}{c}A_L
e^{-a_L T_L} \quad{\rm ``KKLT"}
\\A_L^{(1)} e^{-a_L^{(1)}T_L}+A_L^{(2)} e^{-a_L^{(2)} T_L}\quad {\rm
``KL"}\end{array}\right.\quad. \ee
For simplicity in this section, we will work in units of $M_P$.

Moduli stabilization then proceeds from the F-term scalar potential
for the fields $T_L,T_+,G_-$ which is determined by
\be V_F(T_L,T_+,G_-)=e^K\left(K^{I\bar
J}D_IW\overline{D_JW}-3|W|^2\right) \ee
where $K^{I\bar J}$ is understood to be the inverse K\"ahler metric
derived by keeping the dilaton dependence in $K$ (and thus for its
determination the tree-level dilaton K\"ahler potential
$K_{\tau}=-\ln\left[-i(\tau+\bar\tau)\right]$ has to be included in
$K$). The dilaton is assumed to be fixed by three-form fluxes at
$D_{\tau}W=0$, and we will take $g_s\sim 1/2$ for concreteness.
Thus, here $I,J$ run over the values $L,+,-$, corresponding to the
fields $T_L, T_+, G_-$.

With a choice of parameters in e.g. the KKLT case of
\be A_L=-1\;,\;A_+=1\;,\;a_L=\frac{2\pi}{25}\;,\;a_+={2\pi\over
3}\;,\ W_0=3\times 10^{-2}\quad{\rm ``KKLT"} \ee
%
this  setup stabilizes $T_L\sim 20$, $T_+\sim 4$ and $b\sim 0$ in a
way consistent with the most basic conditions for inflation.
In particular, the moduli potential barriers exceed $V_{inf}$, and
the moduli suffer practically negligible shifts in their VEVs during
inflation driven by an NS5-brane wrapped on the blown-up
two-cycle.~\footnote{Note that given a fully explicit model,
knowledge of the intersection numbers in
$T_+=c_{+\alpha\beta}v^\alpha v^\beta$ may allow for having smaller
values of $T_+$, while still yielding $v_+\sim 2$, as necessary for
sufficiently suppressing the ED1 contributions to the K\"ahler
potential.}

In terms of the supersymmetric multiplets, the NS5-brane potential
is given by
\be\label{VNS5warp} V_{NS5}=M_{\rm
P}^4e^{4A_{bottom}}\frac{(2\pi)^9}{g_s{\cal
V}_E^2}\sqrt{\underbrace{g_s\frac{T_++\bar
T_++\frac{3}{8}g_sc_{+--}(G_-+\bar G_-)^2}{2}}_{{\rm this \; is}
\;v_{+}^2}+g_s^2c^2} \ee
%
with $e^{A_{bottom}}$ denoting the warp factor at the bottom of the
throat.  We obtain a GUT-scale inflaton potential for
$e^{A_{bottom}}\sim 2\times 10^{-4}$. Inputting the axion decay
constant for this case (\ref{warpedsnake}), with $e^{A_{top}}\sim
1$, we find of order $N_w\sim 100$ cycles during inflation, easily
satisfying the backreaction constraint.  Increasing $a_+$ to $\pi$
introduces oscillating corrections to $\eta$ of amplitude $\sim
0.04$.




In all cases discussed above, the uplifting contribution of an
anti-D3-brane \be \delta
V_{\overline{D3}-uplift}=\frac{\delta_{\overline{D3}}}{{\cal
V}_E^{4/3}} \ee is assumed present and is fine-tuned as in
\cite{KKLT}\ so as to provide the post-inflationary minimum at $c=0$
with small positive cosmological constant.


\begin{figure}[t]
\begin{center}
\hspace{-0.5cm}\includegraphics[width=17cm]{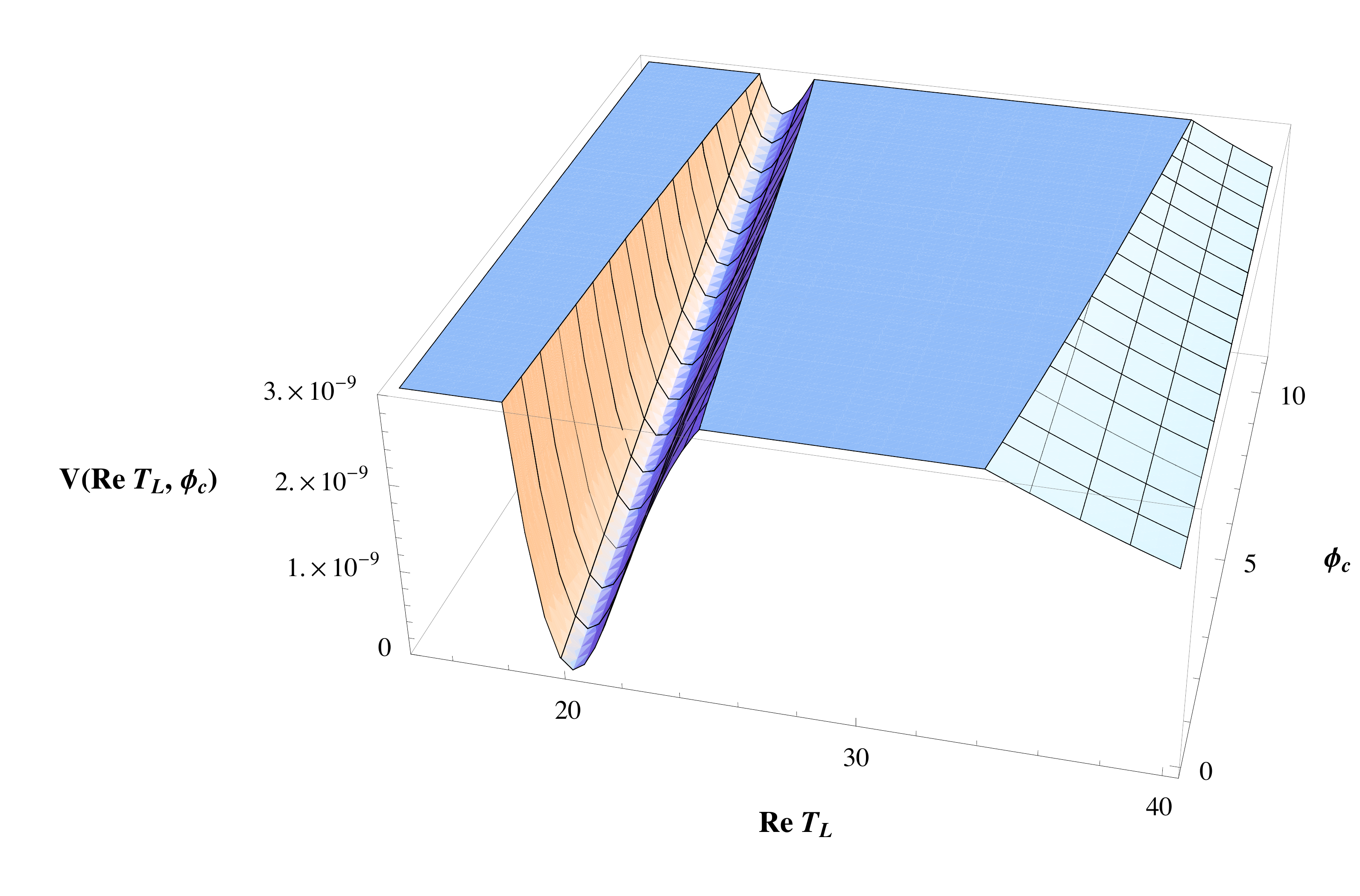}
\end{center}
\refstepcounter{figure}\label{3DPotential}

\vspace*{-.2cm} {\bf Figure~\ref{3DPotential}:} Linear axion
inflaton potential $V({\rm Re}\,T_L,\phi_c)$ with KKLT K\"ahler
moduli stabilization scenario. The linear inflaton valley is clearly
visible. The potential looks very similar (but for the second AdS
minimum at larger volume) for the KL case. The cut-off
surfaces at the top of the plotted box denote the further rise of
the scalar potential in the barriers.
\end{figure}

\subsection{Gravity Waves and Low-Energy Supersymmetry}

It is interesting to consider the possibility of combining
low-energy supersymmetry with high-scale inflation. The present work
moves a step closer to an understanding of this question by
implementing large-field inflation in string compactifications which
have a four-dimensional effective theory with spontaneously broken
${\cal N}=1$ supersymmetry.

In the particular case of KKLT moduli stabilization -- with an
uplift of a SUSY-breaking AdS minimum -- the scale of the moduli
barriers decreases with decreasing scale of supersymmetry breaking.
Kallosh and Linde \cite{KL}\ explained how -- with extra fine-tuning
via an additional racetrack in the superpotential -- one may
decouple these scales (see also the recent work
\cite{Conlon:2008cj}).

In our setup, we may also apply this mechanism, with the following
caveat.  Our wrapped fivebrane action, at its post-inflationary
minimum, itself constitutes a supersymmetry-breaking ``uplifting"
contribution to the potential energy for nonzero $v_+$. This
contribution would need to be very small in order to obtain a low
scale of supersymmetry breaking.  Such a suppression might be
possible by (i) blowing down $v_+$, which may lead to a larger, but
still viable, oscillating contribution to $\eta$ (modulo
suppressions coming from the enhanced local supersymmetry near the
cycle in some examples),
or (ii) warping the NS5-brane further down, as long as this is
consistent with the backreaction constraints.

It is worth emphasizing that despite much progress in recent years,
specific models arise very much under a lamppost, and it is
difficult -- if not impossible -- to determine generic patterns
without a systematic analysis of string
compactifications.\footnote{Moreover, generic compactifications with
this much supersymmetry involve many further ingredients, such as
generalized fluxes, which significantly affect questions of
genericity.}  Thus, although there is no known natural construction
combining high-scale inflation with low scale supersymmetry, neither
is there is a compelling ``no go" theorem.  The answer to this
question must await further development of the subject.

\section{Specific Models II:  Perturbatively Stabilized Compactifications}

Let us next briefly outline some potential examples of our mechanism
in the context of perturbative stabilization of moduli. This class
of examples includes compactifications on more generic --
Ricci-curved -- manifolds, and a correspondingly higher scale of
supersymmetry breaking. The conditions that the flux-induced axion
masses not lift $b$ and $c$, which were automatically satisfied in
the no-scale type IIB Calabi-Yau compactifications discussed above,
will need to be assessed separately in these cases.  The
perturbative models, on the other hand, enjoy some complementary
simplifications of their own, such as the fact that one need not
balance classical effects against nonperturbative effects to
stabilize moduli. The moduli-stabilizing barriers, being power law
in the volume as well as in the dilaton, may be naturally higher,
and the $\eta$ problem for $b$ derived in the previous section does
not directly apply when the volume is perturbatively stabilized.

As with Calabi-Yau compactifications, only a small subset of models
in this class have been analyzed in any detail.  The simplest
examples of this sort involve known classical compactification
geometries and a relatively small set of additional ingredients, and
are therefore accessible to more detailed analysis than the typical
Calabi-Yau compactification (as in
\cite{Silverstein:2007ac,Silverstein:2008sg}).  The most specific,
tractable examples, however, do not incorporate the warping effects
one expects to arise in a typical compactification (whether
low-energy supersymmetric or not).  Clearly the implementation of
our mechanism for linear inflation from axion monodromies will
benefit from further developments in string compactification.

\subsection{Compactifications on Nilmanifolds}

First, consider compactifications of type IIA string theory on a
product of two nilmanifolds
\be \label{nilgeom} ds^2_{Nil\times Nil} =
{L_{u}^2\over\beta}du_1^2+\beta L_{u}^2du_2^2 + L_x^2\left(dx+{M}u_1
d u_2\right)^2  + {L_{u}^2\over\beta}d\tilde u_1^2+\beta
L_{u}^2d\tilde u_2^2 + L_x^2\left(d\tilde x+{M}\tilde u_1 d \tilde
u_2\right)^2, \ee
compactified via projection by a discrete set of isometries, and
stabilized for example with the ingredients described in
\cite{Silverstein:2007ac}, including an orientifold action
exchanging the tilded and untilded coordinates. In the presence of
D4-branes, these manifolds yield monodromy-driven large field
inflation with a $\phi^{2/3}$ potential \cite{Silverstein:2008sg}.
It is interesting to consider the angular closed string moduli in
\cite{Silverstein:2008sg}, to see if monodromy from wrapped branes
might yield linear inflation in axion directions also in these
models.

To begin, we note that the flux couplings in \S3.3\ prevent
inflation in the Neveu-Schwarz axion ($b$) directions in this model,
because of the zero-form flux $m_0$ which plays a leading role in
moduli stabilization. Ramond-Ramond axions come from those
components of $C^{(1)},C^{(3)}$, and $C^{(5)}$ that are invariant
under the orientifold projection. With the NS-NS $H_3$ flux
configuration of the specific example analyzed in
\cite{Silverstein:2007ac}, $C^{(1)}\wedge H_3$ is always nonzero.

Many components of $C^{(3)}$ consistent with the orientifold
projection satisfy $C^{(3)}\wedge H_3 = 0$.  The next question is
whether any ingredients which fit into the compactification
introduce monodromy in one or more of these directions. Consider
(\ref{basicpot}) for the case of an NS5-brane in the presence of a
$C^{(2)}$ axion (i.e. $p=2$). T-duality in a direction $y_\perp$
transverse to the NS5-brane yields a KK5-brane -- a Kaluza-Klein
monopole with fiber direction $y_\perp$.  The T-duality transforms
the $C^{(2)}$ field to a $C^{(3)}$ field with two legs along the
KK5-brane worldvolume and one along $y_\perp$.  Hence a KK5-brane
thus oriented with respect to a $C^{(3)}$ axion $c_3$ introduces a
linear potential for $c_3$.

The setup \cite{Silverstein:2007ac,Silverstein:2008sg} includes of
order $1/\beta$ sets of $M$ KK5-branes wrapped along a linear
combination of the $u_2$ and $\tilde u_2$ directions times a
combination of the $x$ and $\tilde x$ directions, with its fiber
circle in the transverse combination of $x,\tilde x$ directions.
The components of $C^{(3)}$ with legs along these three directions
are lifted by the nilmanifold's ``metric flux" -- that is, the fiber
circle is a torsion cycle.  Thus, in order to implement $c_3$ axion
inflation we need to add additional wrapped branes.

Consider adding a second set of $M$ KK5-branes wrapped along the
$u_2$ and $\tilde u_2$ directions, with their fiber circle in a
linear combination of the $x$ and $\tilde x$ directions.   Let us
denote this set by ${\rm{KK5}}^{\prime}$. They carry a linear
potential in the $c_3$ direction. The ratio of the
${\rm{KK5}}^{\prime}$ potential energy to the original KK5 potential
energy is
\be\label{KKratio} {\beta^{1/2}\over{L_x
L_u}}\sqrt{\beta^2L_u^4+c_3^2g_s^2/L_x^2} \ee
%
We now observe that the decay constant of an axion arising from a
potential that threads a product space of the form
$\Sigma^{(p)}\times\Sigma^{(6-p)}$ is given by
\be \label{productscales}
\phi_c^2\sim {L^{6}\over{(2\pi)^{7}\ell_{(p)}^{2p}\alpha'}}c^2\sim
M_P^2 {g_s^2 c^2\over{2 \ell_{(p)}^{2p}}} ~~~~ {\rm (product~space)}
\ee
Using (\ref{productscales}), (\ref{Nwgeneral}) for $p=3$, we find
that $g_s N_w\sim c g_s/(2\pi)^2\sim 11(2\beta L^3)/(2\pi)^2$. In
order for our added ${\rm{KK5}}^{\prime}$ branes to be subdominant
to the moduli potential all along the inflation trajectory, we need
to tune the anisotropy $\beta$ such that the ratio (\ref{KKratio})
is less than unity.

%
%

%
%

Next let us assess systematically the rest of the consistency
conditions delineated in \S3. First, consider the
${\rm{KK5}}^{\prime}$ branes before the effect of the axion VEV
$c_3$.  The core size of a KK monopole is its fiber size, here
$L_x$; in the present case we obtain $r_{core}^{KK'}\sim M L_x$.
This fits well within the transverse $u_1,\tilde u_1$ directions,
and is marginal for $M\sim 1$ within the transverse linear
combination of $x,\tilde x$ directions.

We now consider the effect of $c_3$ on the core size of the object.
In the present case where our manifold is locally a product space,
the $c_3$ term in the brane action contributes to its effective
tension. In our regime of interest, the tension is of order
${\phi_c\over M_P}\sim 11$ times what its tension would be at
$c_3=0$.  In other words, it behaves like 11 sets of
${\rm{KK5}}^{\prime}$ branes.  This increases the core size by a
factor of 11.

Locally in the $u_1$ directions, the ${\rm{KK5}}^{\prime}$ branes
are BPS objects, and hence the corresponding formula for their
tension applies to good approximation. Moreover, as discussed in
\cite{Silverstein:2007ac}, there are more elaborate methods which
might be used to warp down the tensions of KK5-branes in this space
to separate such marginal ratios of scales, bringing NS5-branes
wrapped on the $x$ and $\tilde x$ directions close to the positions
of the ${\rm{KK5}}^{\prime}$ branes.

Finally, we note that instanton effects which depend on $c_3$ arise
from Euclidean D2-branes.  These are safely suppressed by a factor
of order ${\rm exp}[-\beta L^3/g_s]$.

\subsection{Compactifications on Hyperbolic Spaces (Riemann Surfaces)}

Generic compact manifolds are negatively curved, and moduli
stabilization has been outlined for a very special case of this --
type IIB string theory on a product of three Riemann surfaces
\cite{Riemann}.   Let us therefore sketch the possibilities for
linear inflation from axion monodromies in this class of
compactifications.

Because the volume is directly stabilized, these models do not
suffer from the $\eta$ problem discussed in \S4.1.1 in the
Neveu-Schwarz axion  ($b$) directions.  Moreover, in contrast to the
massive IIA models discussed in the previous subsection, the flux
couplings of \S3.3\ do not immediately lift all the $b$ type axions.
It is therefore possible that inflation with $b$ type axions as well
as $c$ type axions might arise in this example. The main potential
obstruction to this is the rich set of intersecting (p,q)
sevenbranes prescribed in \cite{Riemann}.  Some of these -- in
particular those which combine to form O7-planes -- impose boundary
conditions that components of $B_{MN}$ vanish which are fully
transverse or fully parallel to the O7.  The negative term in the
moduli-fixing potential in \cite{Riemann}\ arises from triple
intersections of (p,q) sevenbranes (which contribute anomalous O3
tension as in \cite{GKP}).

Finally, we note an intriguing feature of more general supercritical
compactifications (of which a special case was studied in
\cite{MSS}) -- compactifications of D-dimensional type II string
theory contain exponentially many RR axions, of order $2^D$ with $D$
the total spacetime dimension.  On the other hand, there are many a
priori possible flux-induced masses for these axions.  Again in this
setting, a systematic analysis of axion monodromy inflation awaits
further progress in the study of string compactification.

\section{Observational Predictions}

We have seen that the monodromy produced by wrapped branes yields a
linear potential, over a super-Planckian distance, for the
canonically-normalized axion field. The leading corrections to this
structure are periodic modulations induced by instantons.

Because of the natural exponential suppression of instanton effects,
it is reasonably straightforward to arrange that these modulations
are negligible, as we argued in our examples above.  When this is
the case, the linear inflaton potential gives for the tensor to
scalar ratio $r$ and tilt $n_s$ of the power spectrum
\be\label{basicnumbers} r\approx 0.07  ~~~~~~~  n_s\approx 0.975  .
\ee
The uncertainty comes only from the usual fact that the number of
e-folds is not known precisely in the absence of a specification of
reheating.  The resulting predictions are indicated in the figure,
which exhibits their consistency with current exclusion contours.
We note that several authors have exhibited a preference in the data \cite{potreconstruction} for potentials with $V''\le 0$, which arises naturally in the case
of monodromy-driven inflation .  Upcoming
CMB experiments promise to reduce these contours to
${\cal{O}}(10^{-2})$ in both directions, which will go a long way
toward discriminating different inflationary mechanisms.

However, it is very interesting to consider the more general case in
which the instanton-induced\footnote{Shifts of the moduli during
inflation may be contrived to give small corrections to the
potential, but this requires inflationary energy that is marginally
sufficient to destabilize the compactification. We expect that most
successful models of inflation based on our mechanism have
negligible corrections from this effect.} modulations of the linear
potential are non-negligible; one example like this might involve a
vanishing $v_+$ in the models described in \S4. In this case we must
incorporate {\it oscillating} corrections to the slow roll
parameters, and correspondingly to the power spectrum.
%
We leave a complete study of this case for the future. For now we
note only that modulations of sufficiently high frequency but
non-negligible amplitude may not affect the {\it average} tilt, but
could conceivably lead to signatures in the more detailed structure
of the power spectrum.
%

\begin{figure}[t]
\begin{center}
\hspace{-0.5cm}\includegraphics[width=17cm]{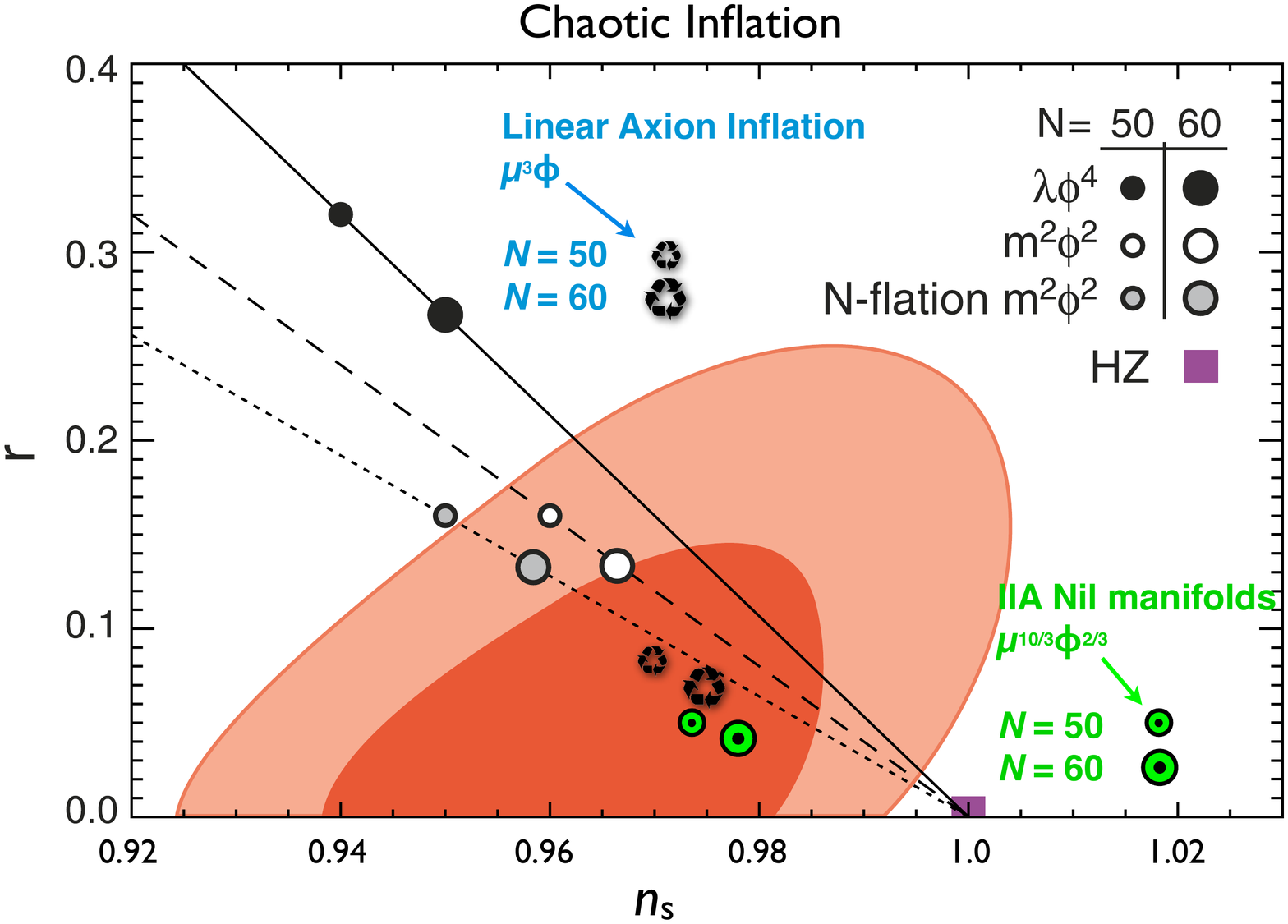}
\end{center}
\refstepcounter{figure}\label{fullpredict}

\vspace*{-.2cm} {\bf Figure~\ref{fullpredict}:} Red: 5-year
WMAP+BAO+SN~\cite{Observations} combined joint 68\% and 95\% error
contours on $(n_s, r)$.
Recycling symbol: general prediction of the linear axion inflation
potential $V(\phi_c)=\mu_{c}^{3}\phi_c$, for $N=50,60$ e-folds before
the end of inflation.
\end{figure}


Let us also remark that the prospect of recurrent modulations of the
perturbation spectrum is quite general in systems making use of the
monodromy mechanism \cite{Silverstein:2008sg}: as the system moves
repeatedly around the monodromy direction, it may interact
periodically with localized degrees of freedom, including, for
example, degrees of freedom into which the system reheats.


\section{Discussion}

Monodromy is a generic phenomenon in string compactifications.  We
have shown that the axion monodromy introduced by space-filling
wrapped fivebranes leads to a linear potential, over a
super-Planckian distance, for the canonically normalized axion
field.  Axion monodromy therefore provides a mechanism for realizing
chaotic inflation, with a linear potential, in string theory.   We
have shown that this mechanism is compatible with various methods of
moduli stabilization, including nonperturbative stabilization of
type IIB string theory on warped Calabi-Yau manifolds, as well as
perturbative stabilization on more general Ricci-curved spaces.
This produces a clear signature in the CMB of $r\approx 0.07$, with
model-dependent opportunities for further, novel, signatures arising
from oscillating corrections to the slow-roll parameters.


Our mechanism is reasonably robust and natural because of the
presence of perturbative axionic shift symmetries.
In our examples, the inflaton potential itself is the leading effect
that breaks the shift symmetry, with instanton corrections naturally
exponentially suppressed. We would like to remark that a related
symmetry structure is plausibly also present in configurations with
more general monodromies not involving axions. Monodromy in the
potential energy arises when a would-be circle in a direction
$\gamma$ in the approximate moduli space is lifted by an additional
ingredient whose potential energy grows as one moves in the $\gamma$
direction.  This unwraps the circle direction and extends the
kinematic range of the corresponding field.  Then, symmetries
translating around the original circle do much to control the
structure of the potential along the eventually-unwrapped direction.
Thus, monodromy-extended directions are not just long; they also
generically profit from approximate symmetries.  Monodromy-extended
directions can be used for large-field inflation if the underlying
moduli potential depends sufficiently weakly on $\gamma$ and if all
corrections to the slow-roll parameters are sufficiently suppressed;
the classes of compactifications analyzed here and in
\cite{Silverstein:2008sg}\ provide two particular realizations of
this effect.\footnote{Recently, monodromy has been used as a method
to model Chain Inflation \cite{ChainInflation}\ in string theory
\cite{chainmonod}.}

There is much more to be done at the level of model building. The
examples we have provided in this work are useful as proofs of
principle, and to that end we have focused on demonstrating
parametric suppression of corrections to the inflaton potential, and
in particular on enumerating a wide array of mechanisms, such as
warping, axionic symmetries, extended local supersymmetry, etc.,
that serve to control such contributions.  We have not yet attempted
to construct a minimal realization of linear axion inflation that
uses the smallest possible subset of these control mechanisms.  This
is an interesting problem for future work, as methods for analyzing
string compactifications and string-theoretic instantons improve.

A further lesson of this work, as of \cite{Silverstein:2008sg}, is
that in large-field models based on monodromy, a degree of
suppression of otherwise problematic contributions to the potential
that suffices for inflation is also sufficient to make firm
predictions for the tilt of the scalar power spectrum.  This is in
sharp contrast to typical small-field models, where fine-tuning the
inflaton potential to be flat enough for inflation is not a strong
enough restriction to be predictive: slight variations in the
fine-tuned contributions can noticeably change the tilt.  The
difference in our case is that the problematic terms arise as
periodic modulations of the potential; requiring that inflation
occurs at all implies that the amplitude of these modulations is
small compared to the scale of changes in the inflaton potential
itself.  In turn, this implies that the {\it average} tilt is not
affected at a detectable level by these modulations.  On the other
hand, it would be very interesting if the oscillations in the
detailed power spectrum produced by a modulated linear potential had
characteristic features accessible to future observations.  In any
case, this class of models is falsifiable on the basis of its
gravity wave signature.

\section*{Acknowledgments}
We thank O. Aharony, T. Banks, D.  Baumann, C. Burgess,  P.
C\'amara, J. Cline,  K. Dasgupta, M. Dine, T. Grimm, S. Kachru, R.
Kallosh, I. Klebanov, A. Linde, J. Maldacena, A. Nicolis, F.
Quevedo, and T. Weigand for useful discussions. We are also grateful
to D. Baumann for assistance with the figures.  L.M. thanks the Stanford
Institute for Theoretical Physics for hospitality during the
completion of this work.  The research of L.M. is supported by NSF
grant PHY-0355005.  The research of E.S. is supported by NSF grant
PHY-0244728, by the DOE under contract DE-AC03-76SF00515, and by BSF
and FQXi grants. The research of A.W. is supported in part by the
Alexander-von-Humboldt foundation, as well as by NSF grant
PHY-0244728.

\begingroup\raggedright\endgroup


\begin{thebibliography}{10}

\bibitem{Inflation}
  A.~H.~Guth,
  ``The Inflationary Universe: A Possible Solution To The Horizon And Flatness
 Problems,''
  Phys.\ Rev.\ D {\bf 23}, 347 (1981);

  A.~D.~Linde,
  ``A New Inflationary Universe Scenario: A Possible Solution Of The Horizon,
  Flatness, Homogeneity, Isotropy And Primordial Monopole Problems,''
  Phys.\ Lett.\ B {\bf 108}, 389 (1982);

  A.~Albrecht and P.~J.~Steinhardt,
  ``Cosmology For Grand Unified Theories With Radiatively Induced Symmetry
  Breaking,''
  Phys.\ Rev.\ Lett.\  {\bf 48}, 1220 (1982).



\bibitem{Linde:1983gd}
  A.~D.~Linde,
  ``Chaotic Inflation,''
  Phys.\ Lett.\  B {\bf 129}, 177 (1983).



\bibitem{Lyth}
  D.~H.~Lyth,
  ``What would we learn by detecting a gravitational wave signal in the  cosmic
  microwave background anisotropy?,''
  Phys.\ Rev.\ Lett.\  {\bf 78}, 1861 (1997)
  [arXiv:hep-ph/9606387].

  G.~Efstathiou and K.~J.~Mack,
  ``The Lyth Bound Revisited,''
  JCAP {\bf 0505}, 008 (2005)
  [arXiv:astro-ph/0503360].


\bibitem{Bmodes}

  M.~Zaldarriaga and U.~Seljak,
  ``An All-Sky Analysis of Polarization in the Microwave Background,''
  Phys.\ Rev.\  D {\bf 55}, 1830 (1997)
  [arXiv:astro-ph/9609170].

  M.~Kamionkowski, A.~Kosowsky and A.~Stebbins,
  ``Statistics of Cosmic Microwave Background Polarization,''
  Phys.\ Rev.\  D {\bf 55}, 7368 (1997)
  [arXiv:astro-ph/9611125].



  G.~Efstathiou,
  ``The Future of Cosmology,''
  arXiv:0712.1513 [astro-ph].

  G.~Efstathiou and S.~Chongchitnan,
  ``The search for primordial tensor modes,''
  Prog.\ Theor.\ Phys.\ Suppl.\  {\bf 163}, 204 (2006)
  [arXiv:astro-ph/0603118].

\bibitem{Bmodeobs}
  A.~C.~Taylor  [the Clover Collaboration],
  ``Clover - A B-mode polarization experiment,''
  New Astron.\ Rev.\  {\bf 50}, 993 (2006)
  [arXiv:astro-ph/0610716].

  C.~J.~MacTavish {\it et al.},
  ``Spider Optimization: Probing the Systematics of a Large Scale B-Mode
  Experiment,''
  arXiv:0710.0375 [astro-ph].

  K.~W.~Yoon {\it et al.},
  ``The Robinson Gravitational Wave Background Telescope (BICEP): a bolometric
  large angular scale CMB polarimeter,''
  arXiv:astro-ph/0606278.

  D.~Samtleben for the QUIET Collaboration,
  ``Measuring the Cosmic Microwave Background Radiation (CMBR) polarization
  with QUIET,''
  arXiv:0802.2657 [astro-ph].

J.~Bock {\it et al.},
  ``Task Force on Cosmic Microwave Background Research,''
  arXiv:astro-ph/0604101.

P.~Oxley {\it et al.},
  ``The EBEX Experiment,''
  Proc.\ SPIE Int.\ Soc.\ Opt.\ Eng.\  {\bf 5543}, 320 (2004)
  [arXiv:astro-ph/0501111].

R.~Charlassier and f.~t.~B.~Collaboration,
  ``The BRAIN experiment, a bolometric interferometer dedicated to the CMB
  B-mode measurement,''
  arXiv:0805.4527 [astro-ph].

Michael J. Myers, William Holzapfel, Adrian T. Lee, Roger O'Brient,
P. L. Richards, Huan T. Tran, Peter Ade, Greg Engargiola, Andy
Smith, and Helmuth Spieler, ``An antenna-coupled bolometer with an
integrated microstrip bandpass filter,"  Appl. Phys. Lett. 86, 114103
(2005).

http://cmbpol.uchicago.edu/

D. Baumann, ...,...,...,..., M. Zaldarriaga, ``Inflationary Observables and Precision Cosmology: Inflation
Working Group White Paper", in preparation.


\bibitem{Freese:1990rb}
  K.~Freese, J.~A.~Frieman and A.~V.~Olinto,
  ``Natural inflation with pseudo - Nambu-Goldstone bosons,''
  Phys.\ Rev.\ Lett.\  {\bf 65}, 3233 (1990).

F.~C.~Adams, J.~R.~Bond, K.~Freese, J.~A.~Frieman and A.~V.~Olinto,
  ``Natural Inflation: Particle Physics Models, Power Law Spectra For Large
  Scale Structure, And Constraints From Cobe,''
  Phys.\ Rev.\  D {\bf 47}, 426 (1993)
  [arXiv:hep-ph/9207245].

\bibitem{OtherNatural}
  N.~Arkani-Hamed, H.~C.~Cheng, P.~Creminelli and L.~Randall,
  ``Extranatural inflation,''
  Phys.\ Rev.\ Lett.\  {\bf 90}, 221302 (2003)
  [arXiv:hep-th/0301218].




\bibitem{Banks:2003sx}
  T.~Banks, M.~Dine, P.~J.~Fox and E.~Gorbatov,
  ``On the possibility of large axion decay constants,''
  JCAP {\bf 0306}, 001 (2003)
  [arXiv:hep-th/0303252].


\bibitem{Nflation}
  S.~Dimopoulos, S.~Kachru, J.~McGreevy and J.~G.~Wacker,
  ``N-flation,''
  arXiv:hep-th/0507205.

  R.~Easther and L.~McAllister,
  ``Random matrices and the spectrum of N-flation,''
  JCAP {\bf 0605}, 018 (2006)
  [arXiv:hep-th/0512102].

  R.~Kallosh, N.~Sivanandam and M.~Soroush,
  ``Axion Inflation and Gravity Waves in String Theory,''
  Phys.\ Rev.\  D {\bf 77}, 043501 (2008)
  [arXiv:0710.3429 [hep-th]].

  T.~W.~Grimm,
  ``Axion Inflation in Type II String Theory,''
  arXiv:0710.3883 [hep-th].

J.~E.~Kim, H.~P.~Nilles and M.~Peloso,
  ``Completing natural inflation,''
  JCAP {\bf 0501}, 005 (2005)
  [arXiv:hep-ph/0409138].



\bibitem{Silverstein:2008sg}
  E.~Silverstein and A.~Westphal,
  ``Monodromy in the CMB: Gravity Waves and String Inflation,''
  arXiv:0803.3085 [hep-th].

\bibitem{Witten:1979ey}
  E.~Witten,
  ``Dyons Of Charge E Theta/2 Pi,''
  Phys.\ Lett.\  B {\bf 86}, 283 (1979).

  N.~Seiberg and E.~Witten,
  ``Electric - magnetic duality, monopole condensation, and confinement in N=2
  supersymmetric Yang-Mills theory,''
  Nucl.\ Phys.\  B {\bf 426}, 19 (1994)
  [Erratum-ibid.\  B {\bf 430}, 485 (1994)]
  [arXiv:hep-th/9407087].

  N.~Seiberg and E.~Witten,
  ``Monopoles, duality and chiral symmetry breaking in N=2 supersymmetric
  QCD,''
  Nucl.\ Phys.\  B {\bf 431}, 484 (1994)
  [arXiv:hep-th/9408099].

\bibitem{wrappedmonod}

See e.g.



P.~S.~Aspinwall and M.~R.~Douglas,
  ``D-brane stability and monodromy,''
  JHEP {\bf 0205}, 031 (2002)
  [arXiv:hep-th/0110071].

  J.~Distler, H.~Jockers and H.~j.~Park,
   ``D-brane monodromies, derived categories and boundary linear sigma
  models,''
  arXiv:hep-th/0206242.



C.~M.~Hull,
  ``A geometry for non-geometric string backgrounds,''
  JHEP {\bf 0510}, 065 (2005)
  [arXiv:hep-th/0406102].


  A.~Lawrence, M.~B.~Schulz and B.~Wecht,
  ``D-branes in nongeometric backgrounds,''
  JHEP {\bf 0607}, 038 (2006)
  [arXiv:hep-th/0602025].

  H.~Jockers,
  ``D-brane monodromies from a matrix-factorization perspective,''
  JHEP {\bf 0702}, 006 (2007)
  [arXiv:hep-th/0612095].


\bibitem{Observations}
  D.~N.~Spergel {\it et al.}  [WMAP Collaboration],
 ``First Year Wilkinson Microwave Anisotropy Probe (WMAP) Observations:
  Determination of Cosmological Parameters,''
  Astrophys.\ J.\ Suppl.\  {\bf 148}, 175 (2003)  [arXiv:astro-ph/0302209];

  H.~V.~Peiris {\it et al.},
 ``First year Wilkinson Microwave Anisotropy Probe (WMAP) observations:
  Implications for inflation,''
  Astrophys.\ J.\ Suppl.\  {\bf 148}, 213 (2003)
  [arXiv:astro-ph/0302225];

  D.~N.~Spergel {\it et al.}  [WMAP Collaboration],
  ``Wilkinson Microwave Anisotropy Probe (WMAP) three year results:
  Implications for cosmology,''
  Astrophys.\ J.\ Suppl.\  {\bf 170}, 377 (2007)
  [arXiv:astro-ph/0603449].

  E.~Komatsu {\it et al.},
  ``Five-Year Wilkinson Microwave Anisotropy Probe (WMAP) Observations:
  Cosmological Interpretation,''
  submitted to Astrophys.\ J.\ Suppl.
  [arXiv:0803.0547 [astro-ph]].

\bibitem{Planck}
  F.~R.~Bouchet  [Planck Collaboration],
  ``The Planck satellite: Status \& perspectives,''
  Mod.\ Phys.\ Lett.\  A {\bf 22}, 1857 (2007).

\bibitem{Svrcek:2006yi}
  P.~Svrcek and E.~Witten,
  ``Axions in string theory,''
  JHEP {\bf 0606}, 051 (2006)
  [arXiv:hep-th/0605206].

\bibitem{Andreas}
  B.~Andreas, G.~Curio and D.~Lust,
  ``The Neveu-Schwarz five-brane and its dual geometries,''
  JHEP {\bf 9810}, 022 (1998)
  [arXiv:hep-th/9807008].

\bibitem{GL}
  T.~W.~Grimm and J.~Louis,
  ``The effective action of N = 1 Calabi-Yau orientifolds,''
  Nucl.\ Phys.\  B {\bf 699}, 387 (2004)
  [arXiv:hep-th/0403067].


\bibitem{Grimm}
  T.~W.~Grimm,
  ``Nonperturbative Corrections and Modularity in N=1 Type IIB
  Compactifications,''
  JHEP {\bf 0710}, 004 (2007)
  [arXiv:0705.3253 [hep-th]].

\bibitem{GKP}
  S.~B.~Giddings, S.~Kachru and J.~Polchinski,
  ``Hierarchies from fluxes in string compactifications,''
  Phys.\ Rev.\  D {\bf 66}, 106006 (2002)
  [arXiv:hep-th/0105097].


\bibitem{Juan}
J. Maldacena, private communication.


\bibitem{KLLS}
  R.~Kallosh, A.~D.~Linde, D.~A.~Linde and L.~Susskind,
  ``Gravity and global symmetries,''
  Phys.\ Rev.\  D {\bf 52}, 912 (1995)
  [arXiv:hep-th/9502069].

\bibitem{RS}
  L.~Randall and R.~Sundrum,
  ``An alternative to compactification,''
  Phys.\ Rev.\ Lett.\  {\bf 83}, 4690 (1999)
  [arXiv:hep-th/9906064].

  L.~Randall and R.~Sundrum,
  ``A large mass hierarchy from a small extra dimension,''
  Phys.\ Rev.\ Lett.\  {\bf 83}, 3370 (1999)
  [arXiv:hep-ph/9905221].

\bibitem{KKLT}
  S.~Kachru, R.~Kallosh, A.~Linde and S.~P.~Trivedi,
  ``de Sitter vacua in string theory,''
  Phys.\ Rev.\  D {\bf 68}, 046005 (2003)
  [arXiv:hep-th/0301240].


\bibitem{LV}
  V.~Balasubramanian, P.~Berglund, J.~P.~Conlon and F.~Quevedo,
  ``Systematics of moduli stabilisation in Calabi-Yau flux
  compactifications,''
  JHEP {\bf 0503}, 007 (2005)
  [arXiv:hep-th/0502058].

  J.~P.~Conlon and F.~Quevedo,
  ``On the explicit construction and statistics of Calabi-Yau flux vacua,''
  JHEP {\bf 0410}, 039 (2004)
  [arXiv:hep-th/0409215].


\bibitem{KKLMMT}
  S.~Kachru, R.~Kallosh, A.~Linde, J.~M.~Maldacena, L.~P.~McAllister and S.~P.~Trivedi,
  ``Towards inflation in string theory,''
  JCAP {\bf 0310}, 013 (2003)
  [arXiv:hep-th/0308055].




\bibitem{KL}
  R.~Kallosh and A.~Linde,
  ``Landscape, the scale of SUSY breaking, and inflation,''
  JHEP {\bf 0412}, 004 (2004)
  [arXiv:hep-th/0411011].


\bibitem{MSS}
  A.~Maloney, E.~Silverstein and A.~Strominger,
  ``De Sitter space in noncritical string theory,''
  arXiv:hep-th/0205316.

E.~Silverstein,
  ``(A)dS backgrounds from asymmetric orientifolds,''
  arXiv:hep-th/0106209.

\bibitem{Riemann}
  A.~Saltman and E.~Silverstein,
  ``A new handle on de Sitter compactifications,''
  JHEP {\bf 0601}, 139 (2006)
  [arXiv:hep-th/0411271].

\bibitem{DGKT}
  O.~DeWolfe, A.~Giryavets, S.~Kachru and W.~Taylor,
  ``Type IIA moduli stabilization,''
  JHEP {\bf 0507}, 066 (2005)
  [arXiv:hep-th/0505160].

\bibitem{Silverstein:2007ac}
  E.~Silverstein,
  ``Simple de Sitter Solutions,''
  Phys.\ Rev.\  D {\bf 77}, 106006 (2008)
  [arXiv:0712.1196 [hep-th]].

\bibitem{KS}
  I.~R.~Klebanov and M.~J.~Strassler,
  ``Supergravity and a confining gauge theory: Duality cascades and
  chiSB-resolution of naked singularities,''
  JHEP {\bf 0008}, 052 (2000)
  [arXiv:hep-th/0007191].

\bibitem{FixingAll}
  F.~Denef, M.~R.~Douglas, B.~Florea, A.~Grassi and S.~Kachru,
  ``Fixing all moduli in a simple F-theory compactification,''
  Adv.\ Theor.\ Math.\ Phys.\  {\bf 9}, 861 (2005)
  [arXiv:hep-th/0503124].

\bibitem{LustII}
  D.~Lust, S.~Reffert, E.~Scheidegger, W.~Schulgin and S.~Stieberger,
  ``Moduli stabilization in type IIB orientifolds. II,''
  Nucl.\ Phys.\  B {\bf 766}, 178 (2007)
  [arXiv:hep-th/0609013].


\bibitem{Witten}
  E.~Witten,
  ``Nonperturbative Superpotentials In String Theory,''
  Nucl.\ Phys.\  B {\bf 474}, 343 (1996)
  [arXiv:hep-th/9604030].

\bibitem{MMMS}
  M.~Marino, R.~Minasian, G.~W.~Moore and A.~Strominger,
  ``Nonlinear instantons from supersymmetric p-branes,''
  JHEP {\bf 0001}, 005 (2000)
  [arXiv:hep-th/9911206].

\bibitem{Intriligator:1995au}
  K.~A.~Intriligator and N.~Seiberg,
  ``Lectures on supersymmetric gauge theories and electric-magnetic  duality,''
  Nucl.\ Phys.\ Proc.\ Suppl.\  {\bf 45BC}, 1 (1996)
  [arXiv:hep-th/9509066].





\bibitem{LustD7with2formflux}
  D.~Lust, S.~Reffert and S.~Stieberger,
  ``Flux-induced soft supersymmetry breaking in chiral type IIb  orientifolds
  with D3/D7-branes,''
  Nucl.\ Phys.\  B {\bf 706}, 3 (2005)
  [arXiv:hep-th/0406092].

\bibitem{Camara}
  P.~G.~Camara and E.~Dudas,
  ``Multi-instanton and string loop corrections in toroidal orbifold models,''
  arXiv:0806.3102 [hep-th].


\bibitem{StringyInstantons}

  R.~Blumenhagen, M.~Cvetic, R.~Richter and T.~Weigand,
  ``Lifting D-Instanton Zero Modes by Recombination and Background Fluxes,''
  JHEP {\bf 0710}, 098 (2007)
  [arXiv:0708.0403 [hep-th]].

\bibitem{SV}

M.~A.~Shifman and A.~I.~Vainshtein,
  ``On holomorphic dependence and infrared effects in supersymmetric gauge
  theories,''
  Nucl.\ Phys.\  B {\bf 359}, 571 (1991).


  N.~Arkani-Hamed and H.~Murayama,
  ``Renormalization group invariance of exact results in supersymmetric  gauge
  theories,''
  Phys.\ Rev.\  D {\bf 57}, 6638 (1998)
  [arXiv:hep-th/9705189].

N.~Arkani-Hamed and H.~Murayama,
  ``Holomorphy, rescaling anomalies and exact beta functions in  supersymmetric
  gauge theories,''
  JHEP {\bf 0006}, 030 (2000)
  [arXiv:hep-th/9707133].


\bibitem{Conlon:2008cj}
  J.~P.~Conlon, R.~Kallosh, A.~Linde and F.~Quevedo,
  ``Volume Modulus Inflation and the Gravitino Mass Problem,''
  arXiv:0806.0809 [hep-th].


\bibitem{potreconstruction}

L.~Alabidi and J.~E.~Lidsey,
  ``Single-Field Inflation After WMAP5,''
  arXiv:0807.2181 [astro-ph].


\bibitem{ChainInflation}
K.~Freese and D.~Spolyar,
  ``Chain inflation: `Bubble bubble toil and trouble',''
  JCAP {\bf 0507}, 007 (2005)
  [arXiv:hep-ph/0412145].

K.~Freese, J.~T.~Liu and D.~Spolyar,
  ``Inflating with the QCD axion,''
  Phys.\ Rev.\  D {\bf 72}, 123521 (2005)
  [arXiv:hep-ph/0502177].

B.~Feldstein and B.~Tweedie,
  ``Density perturbations in chain inflation,''
  JCAP {\bf 0704}, 020 (2007)
  [arXiv:hep-ph/0611286].

Q.~G.~Huang,
  ``Simplified Chain Inflation,''
  JCAP {\bf 0705}, 009 (2007)
  [arXiv:0704.2835 [hep-th]].




\bibitem{chainmonod}
D.~Chialva and U.~H.~Danielsson,
  ``Chain inflation revisited,''
  arXiv:0804.2846 [hep-th].






















\end{thebibliography}
\end{document}